\documentclass[10pt,twocolumn, journal]{IEEEtran}

\usepackage{graphicx,amssymb,lineno}
\usepackage{amsmath,amsfonts,amssymb}

\usepackage{algorithmic}
\usepackage[usenames]{color}
\usepackage{float}

\usepackage[linesnumbered,ruled,vlined]{algorithm2e}

\usepackage{graphicx,graphics,color,epsfig,subfigure,graphpap,rotate}
\usepackage{times, verbatim, subfigure, epsfig, graphicx, latexsym, amsmath}
\usepackage{url}
\usepackage{subfigure}

\usepackage{bm}
\usepackage{cite}

\usepackage{dsfont} %
\usepackage{eqparbox}

\usepackage[marginal]{footmisc}
\usepackage[mathscr]{eucal}
\usepackage{mdwmath}
\usepackage{mdwtab}
\usepackage{multicol}
\usepackage{multirow}%

\usepackage{stmaryrd}
\usepackage{stfloats}

\usepackage{framed}
\definecolor{shadecolor}{rgb}{0.92,0.92,0.92}

\usepackage[amsmath,thmmarks]{ntheorem}
 \usepackage{theorem}  

\newtheorem{Thm}{Theorem}
\newtheorem{Cor}{Corollary}
\newtheorem{lem}{Lemma}
\newtheorem{Def}{Definiton}
\newtheorem{assumption}{Assumption}

\newtheorem{prop}{Proposition}

\newcommand{\mb}[1]{{  \mathbf  #1}}  
\newcommand{\A}{\mb{A}}
\renewcommand{\a}{\mb{a}}
\newcommand{\B}{\mb{B}}
\renewcommand{\b}{\mb{b}}

\renewcommand{\c}{\mb{c}}
\newcommand{\D}{\mb{D}}

\newcommand{\f}{\mb{f}}

\newcommand{\g}{\mb{g}}
\renewcommand{\H}{\mb{H}}
\newcommand{\h}{\mb{h}}
\newcommand{\I}{\mb{I}}

\renewcommand{\P}{\mb{P}}

\newcommand{\Q}{\mb{Q}}

\renewcommand{\S}{\mb{S}}
\newcommand{\s}{\mb{s}}
\newcommand{\T}{\mb{T}}

\newcommand{\U}{\mb{U}}
\renewcommand{\u}{\mb{u}}
\newcommand{\V}{\mb{V}}
\renewcommand{\v}{\mb{v}}
\newcommand{\W}{\mb{W}}
\newcommand{\w}{\mb{w}}
\newcommand{\X}{\mb{X}}
\newcommand{\x}{\mb{x}}
\newcommand{\Y}{\mb{Y}}
\newcommand{\y}{\mb{y}}
\newcommand{\Z}{\mb{Z}}

\newcommand{\THeta}{\mb{\Theta}}

\newcommand{\PHi}{\mb{\Phi}}

\newcommand{\eTa}{\bm{\eta}}

\newcommand{\mU}{\bm{\mu}}

\newcommand{\sIgma}{\bm{\sigma}}

\renewcommand{\vec}[1]{{\mathbf{vec}}\left(#1\right)}

\newcommand{\bcs}{\begin{Cases}}
\newcommand{\ecs}{\end{Cases}}

\def\beq#1\eeq{\begin{equation}#1\end{equation}}
\def\bal#1\eal{\begin{align}#1\end{align}}
\def\bproof#1\eproof{\begin{IEEEproof}#1\end{IEEEproof}}

\def\blem#1\elem{\begin{lem}#1\end{lem}}
\def\bthm#1\ethm{\begin{Thm}#1\end{Thm}}
\def\bcor#1\ecor{\begin{Cor}#1\end{Cor}}
\def\bdef#1\edef{\begin{Def}#1\end{Def}}
\def\bprop#1\eprop{\begin{prop}#1\end{prop}}
\def\bremark#1\eremark{\begin{Remark}#1\end{Remark}}
\def\bassum#1\eassum{\begin{assumption}#1\end{assumption}}

\begin{document}
 \markboth{IEEE TRANSACTlONS  ON  Signal Process, xx  }{ panjian: \paperTitleMarkboth} 

\title{ Fast Two-Dimensional Atomic Norm Minimization in Spectrum Estimation and Denoising }
\author{Jian~Pan \IEEEmembership{Student~Member,~IEEE}, Jun~Tang, \IEEEmembership{Member,~IEEE}, and Yong Niu, ~\IEEEmembership{Member,~IEEE}
\thanks{Jian~Pan, and Jun~Tang are with the Department of Electronic Engineering,
        Tsinghua University, Beijing, 100084, China (email: jianlonger@126.com, tangj\_ee@mail.tsinghua.edu.cn). Yong Niu is with State Key Laboratory of Rail Traffic Control and Safety, Beijing Jiaotong University, Beijing 100044, China (E-mails: niuy11@163.com). This work was supported in part by National Natural Science Foundation of China under Grant 61171120.}}

\maketitle
\IEEEpeerreviewmaketitle

\vspace{-1.5cm}

\begin{abstract} 
Motivated by recent work on two dimensional (2D) harmonic component recovery via atomic norm minimization (ANM), a fast 2D direction of arrival (DOA) off-grid estimation based on ANM method was proposed. By introducing a  matrix atomic norm the 2D DOA estimation problem is turned into matrix atomic norm minimization (MANM) problem. Since the 2D-ANM gridless DOA estimation is processed by vectorizing the 2D into 1D estimation and solved via semi-definite programming (SDP), which is with high computational cost in 2D processing when the number of antennas increases to large size. In order to overcome this difficulty, a detail formulation of MANM problem via SDP method is offered in this paper, the MANM method converts the original $MN+1$ dimensions problem into a $M+N$ dimensions SDP problem and greatly reduces the computational complexity. In this paper we study the problem of 2D line spectrally-sparse signal recovery from partial noiseless observations and full noisy observations, both of which can be solved efficiently via MANM method and obtain high accuracy estimation of the  true 2D angles. We give a sufficient condition of the optimality condition of the proposed method and prove an up bound of the expected error rate for denoising. Finally, numerical simulations are conducted to show the  efficiency and performance of the proposed method, with comparisons against several existed sparse methods.
\end{abstract}



\section{Introduction}
 Estimating the frequencies of a superposition of complex harmonic 2D signals arises in  many applications, including the direction of arrival
estimation in radar target tracking and location \cite{radar1}\cite{radar2}, multiple-input multiple-output (MIMO) antennas communication channel estimation \cite{cm1}\cite{cm2}, and super-resolution imaging through a Fourier imaging system \cite{img}. All of these applications can be attributed to 2D DOA estimation problem.

Conventional 2D DOA estimation methods are often based on Subspace theory methods such as 2D unitary ESPRIT \cite{ESPRIT} , 2D MUSIC \cite{MUSIC}, and the Matrix Enhancement Matrix Pencil (MEMP) method \cite{MEMP}, etc. However, most of these methods rely on the covariance of sampled observation, which requires that the number of independent observations should  be larger than the antenna numbers. what's more, they may fail for coherent sources \cite{coh}.
Since last decade, researchers have put forward super resolution DOA estimation methods based on compressed sensing (CS) \cite{CS} theory, which exploits source sparsity for frequency or angular estimation \cite{CS2}\cite{CS3} and enables DOA estimation even from a single snapshot of measurements, regardless of source correlation. The CS based DOA estimation methods are implemented  via discretizing the 2D possible angle scope with grid points and then recovering the true location by sparse recovery methods such as basis persuit (BP)\cite{BP}, least absolute shrinkage and selection operation (LASSO) \cite{LASSO}, orthogonal matching persuit (OMP) \cite{OMP} and so on. These sparse methods result in fine estimation performance if the true angle location satisfies the pairwise isometry property (PIP) \cite{PIP} and the grids are dense enough.  However, researchers have found that if the observed signal contains off-grid harmonic components it may result in considerable performance decreasing via conventional sparse approximation algorithms over the discrete basis, which has been pointed out in \cite{miss1}\cite{miss2} \cite{miss3}.
Therefore, it is necessary to design a parameter estimation method, which is not relevant to the priori basis for sparse reconstruction while still utilizing the sparsity property.

Recently, a new gridless sparse recovery method for spectral estimation is developed via atomic norm minimization (ANM) \cite{Tang}, which provides a structure-based optimization approach that utilizing the Vandermonde structure of the captured data in a SDP via Toeplitz matrix \cite{Carath1} \cite{Carath2}. The ANM method works well in recovering a spectrally-sparse signal from a few numbers of randomly selected samples of the complete observation data\cite{Tang2}\cite{super}. However, ANM method can not be directly extended to the 2D case by vectorizing  the matrix data into vector data and solving the 1D ANM optimum problem, i.e. the equivalence between the 2D atomic norm minimization in its 1D SDP form is not guaranteed, since the Vandermonde decomposition of Toeplitz matrix via Caratheodory's theorem   does not hold in higher dimensions. Poineered by Chi \cite{Chi} vectorized -ANM for 2D frequency estimation has been exploited, which turns the 2D frequency estimation into a double-folds Toeplitz matrix SDP problem. This method retains the merits of the ANM approach in terms of super-resolution from single-snapshot observations, and robustness to source correlation. But its computation complexity is very high, which becomes almost intractable when the dimensions of antenna array become large.

In this paper, a new definition of ANM is formulated via introducing a new atom set based on 2D harmonic component matrix which turns the double-folds Toeplitz matrix into two Toeplitz matrices with harmonic component contained in one dimension. Accordingly, a new SDP problem is formed for the matrix ANM (MANM), which has greatly decreased the optimum problem's size and remarkably improved computational efficiency. The run time of MANM method is several orders of magnitude  lower than of VANM method, while all other advantages  of ANM are reserved such as high accuracy, high resolution and a small number of observation requirements. Theoretic formulations of our method are given in both  proofs and simulations to validate the proposed MANM method.

The rest of this  paper is organized as follows. In Section II, system model and problem,  related literature and introduction to MANM are described. In Section III, the proposed matrix atomic norm minimization algorithm for full data recovery from partial noiseless observations and data denoise from full noisy observations via SDP
are formulated, and their performance guarantee and proof are deferred
to the Appendix B and Appendix C. In Section IV, Numerical simulations are conducted to test the efficiency and performance of proposed algorithm. Finally, this paper is concluded in Section V.

\emph{Notations:} Throughout the paper, matrices are denoted by bold capital letters, and vectors by bold lowercase letters. $(\cdot)^T, (\cdot)^{*}, (\cdot)^H$ denote the transpose, conjugate and conjugate transpose, respectively. $\|\cdot\|$ denotes the Euclidean norm and  $\|\cdot\|_F$ denotes the Frobenius norm of matrix. $\mathrm{Tr}(\cdot)$ denotes the trace of a matrix, and $\langle \A,\B \rangle \triangleq \mathrm{Tr}(\B^H\A)$ stands for the inner product of matrix $\A$ and $\B$.  $\langle \A,\B \rangle_\mathds{R} \triangleq \mathrm{Re}(\langle \A,\B \rangle )$ represent the real part of the inner product. $\otimes$ denotes the Kronecker product. $\mathds{R}$ and $\mathds{C}$ denotes the sets of all real numbers and complex numbers, respectively. $\mathds{P}(\cdot)$ denote the probability of incident $(\cdot)$ and $\mathds{E}(\cdot)$ denotes the expectation of the argument.

\section{Problem Formulation and the definition of atomic norm}
In this section the sparse 2D-harmonic signal model is formulated via a typical example and a new kind of atomic norm is introduced to fast estimate 2D harmonic component.
\subsection{Signal Model}
\begin{figure}[!htbp]
      \centering
       \includegraphics[width=6cm]{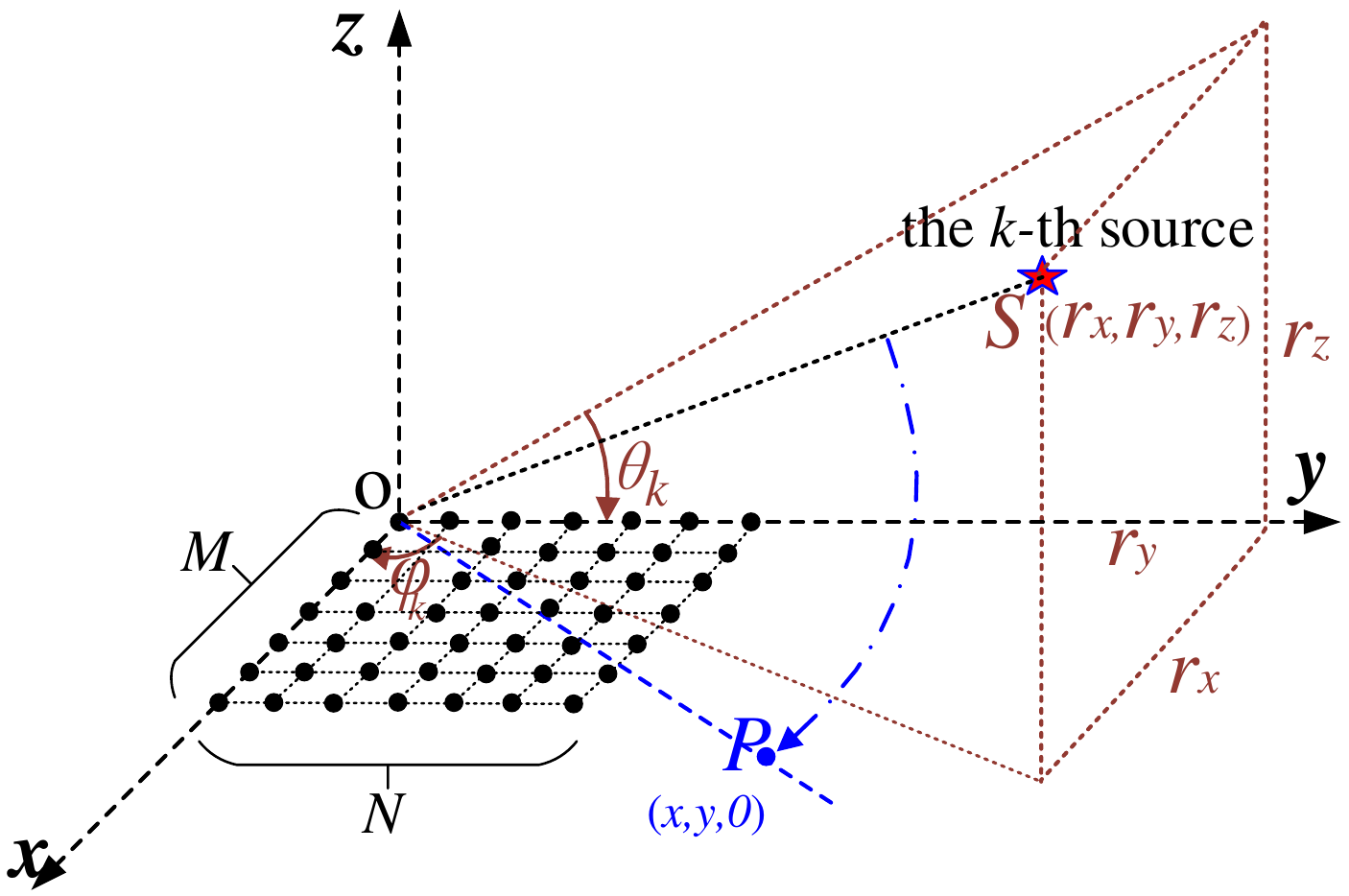}
    \caption{ Source signal in far field for a $M \times N $ Uniform Rectangle Array (URA)}
    \label{fig_1}
\end{figure}

Assume that there are $K$ source signals in far-field scene with their echo impinge on an  $M\times N$ Uniform Rectangle Array (URA) which contains $M$ and $N$ antenna elements along  $x$-direction and $y$-direction. Let $(\theta_k,\varphi_k)$ be the azimuth angle  and elevation angle of the  $k$-th source signal as is shown in Fig. \ref{fig_1}. \\
As for the $k$-th source signal, let $(r_x,r_y,r_z)$ denotes the location of the $k$-th source signal in 3 dimension coordinates, then for a point $P$ locate at $(x,y,0)$  in $x-O-y$ plane, the path difference with respect to the antenna element at origin $O$ (coordinate $(0,0,0)$)  is
\beq
\Delta d = \frac{r_x \cdot x + r_y\cdot y}{\sqrt{r_x^2+r_y^2+r_z^2}}\quad\quad s.t.
\begin{cases}
\tan(\theta_k)=r_z/\sqrt{r_x^2+r_y^2} \\
 \tan(\varphi_k)=r_y/r_x.
 \end{cases}
\eeq
Let $d_x,d_y$ denote the inter-element spacing with respect to $x$-direction and $y$-direction, if $(x,y)=(md_x,nd_y)$ then the phase difference $\Delta \Phi$ that with respect to the antenna element at origin $O$ is
\begin{align}
\Delta \Phi &= 2\pi m\cdot f_{x,k} + 2\pi n\cdot f_{y,k}
\\
 f_{x,k}&\triangleq \frac{1}{\lambda_c}\cos\varphi_k \cos\theta_k \label{a1}\\
  f_{y,k}&\triangleq \frac{1}{\lambda_c}\sin\varphi_k \cos\theta_k  \label{a2}
\end{align}
 where $\lambda_c$ is the wavelength of received echo.

 Hence, the total received target echo data matrix without noise to this URA can be written as
  \begin{align}
  \X &=[x_{m,n}]_{M\times N}    \quad _{0\leq m\leq M-1,0\leq n \leq N-1}\notag\\
  &x_{m,n}=\sum_{k=0}^{K}\frac{1}{\sqrt{MN}}s^{\star}_k\cdot e^{-j2\pi\cdot( m f_{x,k}+ n f_{y,k})}\label{Xdata}
  \end{align}
  where, $s^{\star}_k \in \mathds{C}$ denotes the echo strength of the $k$-th source signal, $(f_{x,k}, f_{y,k})$ are normalized spatial frequency  factors that decided by $(\theta_k,\varphi_k)$ via (\ref{a1}) and (\ref{a2}). It is easy to see  that (\ref{Xdata}) can be rewritten in a brief form as
  \begin{align}
  \X=\sum_{k=0}^{K}s^{\star}_k\cdot\a_{_X}( f_{x,k}) \cdot \a_{_Y}^T( f_{y,k})=\A_{_X}\cdot\S\cdot\A_{_Y}^T   \label{Xdata2}
  \end{align}
  where
  \begin{align}
  \a_{_X}(f_{x,k})&\triangleq [1,e^{-j2\pi\cdot f_{x,k}},...,e^{-j2\pi\cdot(M-1) f_{x,k}}]^T\\
  \a_{_Y}(f_{y,k})&\triangleq [1,e^{-j2\pi\cdot f_{y,k}},...,e^{-j2\pi\cdot(N-1)  f_{y,k}}]^T \\
  \S&\triangleq \textbf{diag}(\s^{\star}), ~ \s^{\star}=[s^{\star}_1,s^{\star}_2,...,s^{\star}_K]^T\\
  \A_{_X}&\triangleq [\a_{_X}(f_{x,1}),\a_{_X}(f_{x,2}),...,\a_{_X}( f_{x,K})]\\
  \A_{_Y}&\triangleq[\a_{_Y}(f_{y,1}),\a_{_Y}(f_{y,2}),...,\a_{_Y}( f_{y,K})].
  \end{align}

  The goal of 2D DOA estimation for URA is identical to recovery $\{( f_{{_{x,k}}}, f_{_{y,k}})\}_{k=1}^K$  from its observation $\X$ in (\ref{Xdata}), since  $\{(\theta_k,\varphi_k)\}_{k=1}^K$ can be obtained by solving (\ref{a1}) and (\ref{a2}). without loss of generality, we turn to estimate   $\{( f_{{_{x,k}}}, f_{_{y,k}})\}_{k=1}^K$  instead.

  \subsection{Vectorial Atomic Norm for 2D DOA model}
  In \cite{Chi}\cite{VD} the Vectorial Atomic Norm is defined via vectorized the data matrix to get an large dimension atom set. Let $\x=\vec{\X}\in \mathds{C}^{MN\times MN}$ be the vectorized data matrice, then one has
  \begin{align}
  \x & =(\A_{_Y}\otimes\A_{_X})\cdot\vec{\S}=\sum_{k=1}^Ks^{\star}_k\cdot\a_{_Y}(f_{y,k}) \otimes \a_{_X}(f_{x,k}) \notag \\
  &=\sum_{k=1}^K \sqrt{MN}s^{\star}_k\cdot\c(\f_k)
  \end{align}
  where
  \beq
  \c(\f_k) \triangleq  \c(f_{_{x,k}},f_{_{y,k}})= \frac{1}{\sqrt{MN}} \a_{_X}(f_{x,k}) \otimes \a_{_Y}(f_{y,k}). \label{cc1}
  \eeq

  The atom set $\mathcal{A}_v$ is defined as the collection of all normalized 2-D complex harmonic:
  \begin{align}
  \mathcal{A}_v = \left\{ \c{(\f)} ~\big{|}~ \c(\f)\in \mathds{C}^{MN},\f\in[0,1]\times [0,1]\right\} \notag \\
  \c(\f) \triangleq  \c(f_{_{x}},f_{_{y}})= \frac{1}{\sqrt{MN}} \a_{_X}(f_{x}) \otimes \a_{_Y}(f_{y})
  \end{align}
  and the atomic norm of $\x$ can be caculated as
  \begin{align}
  \|\x\|_{\mathcal{A}_v} &=  {\mathrm{inf}} \Big{\{} \sum_k|d_k|  ~\big{|}~  \x=\sum_kd_k\c(\f_k),\c(\f_k)\in\mathcal{A}_v \Big{\}} \notag \\
  &=\underset{\U\in\mathds{C}^{M\times N}\atop v\in \mathds{C}}{\mathrm{min}} \quad \Big{\{}\frac{1}{2}v+  \frac{1}{2}\mathrm{Tr} \Big{(}\mathcal{T}_{2D}(\U)\Big{)}      \Big{\}} \notag \\
  & \quad \quad s.t \left[
          \begin{array}{cc}
            \mathcal{T}_{2D}(\U) & \x \\
            \x^H & v \\
          \end{array}
        \right] \succeq \mathbf{0} \label{VANM}
  \end{align}
where,  $\U=[u_{(i,k)}]_{M\times N }$  and $\mathcal{T}_{2D}(\U)$ is a two-fold block Hermite Toeplitz matrix which is defined as
\begin{align}
\mathcal{T}_{2D}(\U)& =
              \left[
                  \begin{array}{cccc}
                  \T_1 & \T_2 & ... & \T_{M}\\
                  \T_2^* & \T_1 & ... & \T_{M-1}\\
                         : & : & ... & : \\
                  \T_{M}^* & \T_{M-1}^* & ... & \T_{1}\\
                       \end{array}
                     \right], \label{T1} \\
    \T_{l} &=  \left[
              \begin{array}{cccc}
               u_{(l,1)} & u_{(l,2)} & ... & u_{(l,N)}\\
               u_{(l,2)}^* & u_{(l,1)} & ... & u_{(l,N-1)}\\
                : & : & ... & : \\
               u_{(l,N)}^* & u_{(l,N-1)}^*  & ... & u_{(l,1)}\\
              \end{array}
         \right].  \label{T2}
\end{align}
\subsection{Matrix Atomic Norm for 2D DOA model}
  According to the signal model (\ref{Xdata2}) to (11), the matrix atom set is defined as all matrices  that decided by $\{(f_{_{x,k}},f_{_{y,k}})\}_k$, i.e.,
  \begin{align}
   \mathcal{A}_m & \triangleq \left\{ \A_m(\f)  ~\big{|}~  \f \in[0,1]\times [0,1]\right\} \label{Am}
  \end{align}
  where
   \begin{align}
   \A_m(\f_k) = \a_{_{_X}(_{f_{x,k}})} \cdot \a_{{_Y}(f_{y,k})}^T \quad(f_{_{x,k}},f_{_{y,k}})\in[0,1]\times [0,1] \notag
  \end{align}
  and  the atomic norm of $\X$ can be caculated as
  \begin{align}
  \|\X\|_{\mathcal{A}_m} &=  {\mathrm{inf}} \big{\{} \sum_k|s_k|  ~\big{|}~  \X=\sum_k s_k\A_{m(\f_k)}, \A_{m(\f)}\in\mathcal{A}_m \big{\}}. \label{XAd}
  \end{align}
 which is the gauge function associated with the convex hull
of $\mathcal{A}_m$. Encouragingly, the matrix atomic norm satisfies the following
equivalent SDP form, which can be computed efficiently.
The proof can be found in Appendix A.
\begin{Thm}\label{Thm1}
If $\X \in \mathds{C}^{M \times N}$, then the matrix atom norm $\|\X\|_{\mathcal{A}_m}$ can be rewritten as
\begin{align}
\|\X\|_{\mathcal{A}_m} &=\underset{(\u,\v)\in\mathds{C}^M \times \mathds{C}^N}{\mathrm{inf}} \quad{\mathrm{Tr}}  \Big{(} {\frac{1}{2M}\mathcal{T}(\u)+  \frac{1}{2N}\mathcal{T}(\v)} \Big{)} \notag \\
  & \quad \quad s.t. \left[
          \begin{array}{cc}
            \mathcal{T}(\u) & \X \\
            \X^H & \mathcal{T}(\v) \\
          \end{array}
        \right] \succeq \mathbf{0} \label{atomeq}
\end{align}
where, $ \mathcal{T}(\u) $ and $\mathcal{T}(\v) $ are Hermite Toeplitz matrices with vector $\u$ and $\v$ as their first column respectively.
\end{Thm}
Meanwhile, in some cases, it is  helpful to analysis and implement some algorithm induced by norm minimization via introducing the definition of dual norm of $\|\X\|_{\mathcal{A}_m}$ , which is defined as
\begin{align}
\|\Z\|_{\mathcal{A}^{'}_m}&= \underset{\|\X\|_{\mathcal{A}_m}\leq 1}{\mathrm{sup}}  \langle  \Z,\X\rangle_{\mathds{R}} \notag \\ &= \underset{\f_k \in[0,1]\times [0,1]\atop \sum_k|s_k|\leq 1}{\mathrm{sup}}  \langle  \Z,\sum_{k}s_k\A_d(\f_k)\rangle_{\mathds{R}} \notag \\
&=\underset{\f \in[0,1]\times [0,1]}{\mathrm{sup}}   \langle  \Z,\A_m(\f)\rangle_{\mathds{R}}
 \label{dual1}
\end{align}
Here, $(\mathcal{A}^{'}_m,~\|\cdot\|_{\mathcal{A}^{'}_m})$ and $(\mathcal{A}_m,~\|\cdot\|_{\mathcal{A}_m})$ are dual normal space to each other.  By weak duality, it always holds that
\beq
|\langle\X,\Z\rangle| \leq \|\Z\|_{\mathcal{A}^{'}_m}\cdot\|\X\|_{\mathcal{A}_m}  \label{dual2}
\eeq

\section{MANM Algorithm For 2D DOA}
In this section, we consider two issues about 2D spectrum estimation and denoising
via matrix  atomic norm minimization from partial or
noisy observations, i.e, (a) recovering the original completed data $\X^{\sharp}$ from their partial noiseless observations $\X_{\Omega}$; and (b) denoising
from their full observations $\X$ in AWGN model.
\subsection{Signal Recovery From Partial Noiseless Observations}
Assume that a random or deterministic (sub)set of entries
of data $\X_{\Omega}$ defined in (\ref{Xdata}) and (\ref{Xdata2}) are observed, and the index set of observation entries is denoted by $\Omega\subset\{1,2,..M\}\times\{1,2,...,N\}$. In the noise free case, the real complete data matrix $\X^{\sharp}$ can be  recovered from its partially observed entries via MANM problem as following
\beq
\X=\underset{\X}{\mathrm{argmin}}  \|\X\|_{\mathcal{A}_m}\quad s.t.\quad \X_{\Omega}=\X^{\sharp}_{\Omega}. \label{DANMnf}
\eeq
where, $\X_{\Omega}$ denotes the entries of $\X$ whose index are in set $\Omega$.  According to theorem \ref{Thm1} optimum problem (\ref{DANMnf}) can be solved via following semidefinite program,
\begin{align}
\underset{(\u,\v,\X)}{\mathrm{argmin}}~\quad~&{\mathrm{Tr}}  \Big{(} {\frac{1}{2M}\mathcal{T}(\u)+  \frac{1}{2N}\mathcal{T}(\v)} \Big{)}  \notag \\
  s.t. & \quad \left[
          \begin{array}{cc}
            \mathcal{T}(\u) & \X \\
            \X^H & \mathcal{T}(\v) \\
          \end{array}
        \right] \succeq \mathbf{0}  \notag \\
        &\quad ~~ \X_{\Omega} = \X^{\sharp}_{\Omega}. \label{optalg1}
\end{align}
It ia easy to solve this SDP problem by using CVX soft parcel \cite{cvx}. While, in order to seek the uniqueness  condition for solution of problem (\ref{DANMnf}), we resort to analysis its dual problem.  According to (\ref{dual1}) it can be obtained that the dual problem of (\ref{DANMnf}) is
\beq
\underset{\Z}{\mathrm{max}}  |\langle\Z,\X^\sharp\rangle|  \quad s.t. \quad  \|\Z\|_{\mathcal{A}^{'}_m}\leq 1~,~ \Z_{\Omega^c}=\mathbf{0} \label{DP1}
\eeq
where $\Omega^c \triangleq \{1,2...,M\}\times \{ 1,2,...,N\} \setminus \Omega$ denotes the complementary index set of $\Omega$.

Let $(\X,\Z)$ be primal-dual feasible solution to (\ref{DANMnf})  and (\ref{DP1}), then $|\langle\Z,\X\rangle|=|\langle\Z,\X^\sharp\rangle|=\|\X^\sharp\|_{\mathcal{A}_m}$
holds if and only if $\Z$ is dual optimal and $\X$
is primal optimal. Utilizing strong duality, we have the following
proposition to certify the optimality of the solution of (\ref{DANMnf}).
\begin{prop}
The solution of optimum problem (\ref{DANMnf}) is unique and equal to the  complete data $\X^\sharp=\sum_{j\in J } s_j\A_m(\f_j)$  if there exists $\Z$ and $ Q(\f)\triangleq \langle \Z,\A_m(\f)\rangle$ that
satisfying
\begin{align}
\begin{cases}
 \Z_{\Omega^c} =\mathbf{0} \\
 Q(\f_j) ={\mathrm{sign}}(s_j),  \quad \forall \f_j\in \mathcal{F} \\
 |Q(\f)| < 1 , \quad\quad\quad\quad \forall \f \notin \mathcal{F}
\end{cases}
\label{uc1}
\end{align}
where, $ \mathcal{F}=\{\f_j\}_{j\in J }$ which including all 2D-frequency components of $\X^\sharp$.
\end{prop}
\begin{IEEEproof}
First, any $\Z$ satisfying (\ref{uc1}) is dual feasible. one has $\|\Z\|_{\mathcal{A}^{'}_m} \leq 1$ and
\begin{align}
\|\X^\sharp\|_{\mathcal{A}_m} &\geq \|\X^\sharp\|_{\mathcal{A}_m}  \|\Z\|_{\mathcal{A}^{'}_m} \geq
| \langle  \Z,\X^\sharp \rangle|\notag\\
 &=|\langle  \Z,\sum_{j\in J } s_j\A_m(\f_j) \rangle|
 =|\sum_{j\in J } s_j^* \langle  \Z,\A_m(\f_j) \rangle|\notag\\
 &=|\sum_{j\in J }s_j^* Q(\f_j)|
=\sum_{j\in J } |s_j| \geq \|\X^\sharp\|_{\mathcal{A}_m}.
\end{align}
Hence $|\langle  \Z,\X^\sharp \rangle|=\|\X^\sharp\|_{\mathcal{A}_m}$. By strong duality it holds that $\X^\sharp$ is primal optimal and $\Z$ is dual optimal.

For uniqueness, assume that $\hat{\X}= \sum_{j} \hat{s}_j\A_m(\hat{\f}_j) $ with $\|\hat{\X}\|_{\mathcal{A}_m}=\sum_j |\hat{s}_j|$ is another optimal solution. Then one has
\begin{align}
|\langle  \Z,\hat{\X} \rangle|
&=|\langle  \Z,\sum_{j} \hat{s}_j\A_m(\hat{\f}_j) \rangle|
 \notag\\
 &=|\sum_{\f_j\in \mathcal{F} } \hat{s}^*_j\langle  \Z,  \A_m(\f _j) \rangle + \sum_{\hat{\f}_l\notin \mathcal{F} }\hat{s}^*_l\langle  \Z,  \A_m(\hat{\f} _l) \rangle |
 \notag \\
  & \leq \sum_{\hat{\f}_j\in \mathcal{F}} |\hat{s}_j| + \sum_{\hat{\f}_l\notin \mathcal{F} }|\hat{s}_l|  = \|\hat{\X}\|_{\mathcal{A}_m} \notag
\end{align}
which contradicts strong duality. Therefore the optimal solution
of (\ref{DANMnf}) is unique.
\end{IEEEproof}

Proposition 1 is a sufficient condition other than necessary condition to guarantee  the uniqueness of optimality solution for problem (\ref{DANMnf}), i.e., as long as we can find a dual polynomial that satisfies (\ref{uc1}), then $\X=\X^\sharp$  is the unique optimum solution of problem (\ref{DANMnf}). while, the inverse claims not necessarily hold.

{\bf{Remark 1: Reducing Complexity:}} The MANM-SDP (\ref{opt1}) is to solving SDP problem from a constraint of size $(M+N)\times (M+N)$, while  as for  the vectorial ANM-SDP (\ref{VANM}) the constraint size is $(MN+1)\times (MN+1)$. Hence the MANM method is able to remarkably decrease the constraint size in SDP especially when the dimension of array ($M,N$)  are very large. This also can be seen in their time consuming level in  simulation results in section \ref{sim}.

{\bf{Remark 2: 2D-Frequencies estimation:}}
It is often the case that the harmonic components $\f$ are more focused than the completeness of the observation data matrix $\X$. while  the MANM-SDP algorithm (\ref{opt1}) offers us a way to directly recover the harmonic components information of $\f^{\sharp}=\{(f_{x,i}^{\sharp},f_{y,i}^{\sharp})\}_{i=1,..K}$  from  $\mathcal{T}(\u)$ and $\mathcal{T}(\v)$ via
\beq
\begin{cases}
\mathcal{T}(\u)=\A_{_X(\f_x^{\sharp})}\cdot\D_x\cdot\A_{_X(\f_x^{\sharp})}^T,    \quad  \D_x \succ \mathbf{0} \\
\mathcal{T}(\v)=\A_{_Y(\f_y^{\sharp})}\cdot\D_y\cdot\A_{_Y(\f_y^{\sharp})}^T,    \quad   \D_y \succ \mathbf{0}.
\end{cases}
\label{theta1D}
\eeq

{\bf{Remark 3: 2D-Frequencies Pairing:}} By solving the SDP in (\ref{opt1} ), the 2D harmonic information can be obtained  via Vandermonde decomposition on their one-level Toeplitz matrices, which is much simpler than the two-level decomposition in [10], [11], and then pairing the harmonic components into $K$ pairs by maximum correlation method as following,
\begin{align}
j_i=\underset{j}{\mathrm{argmax}}\quad  |\a_{_X(f_{x,i})}^H \cdot \hat{\X}\cdot\a_{_Y(f_{y,j})}^{*}|. \label{pairing}
\end{align}
where, $j_i$ denotes the frequency index of $\f_y$ that matched to $f_{x,i}$, and $\hat{\X}$ denotes the recovered data matrix.
\subsection{Signal Denoise From Full Observed Data }
In this section, we consider the problem of 2D DOA  in additive noise case when full observations are available with theirs observation model given as
\beq
\Y=\X^\sharp + \W    \label{noimod}
\eeq
where, $\W=[w_{i,j}]_{M\times N}\in\mathds{C}^{M\times N}$ is the noise data part,  $\X^\sharp $ is the real complete data part. In this case, the matrix atomic norm regularized recovery method is proposed as
\beq
\hat{\X}= \underset{\X}{\mathrm{argmin}} \frac{1}{2}\|\Y-\X\|^2_F+\lambda\|\X\|_{\mathcal{A}_m} \label{denoise}
\eeq
here, $\lambda$ is a positive regularization parameter. It is easy to see that (\ref{denoise}) is equivalent to seek the following SDP problem
\begin{align}
\underset{(\u,\v,\X)}{\mathrm{argmin}}~~&\frac{1}{2}\|\Y-\X\|^2_F+\lambda{\mathrm{Tr}}  \Big{(} {\frac{1}{2M}\mathcal{T}(\u)+  \frac{1}{2N}\mathcal{T}(\v)} \Big{)}  \notag \\
  s.t. & \quad \left[
          \begin{array}{cc}
            \mathcal{T}(\u) & \X \\
            \X^H & \mathcal{T}(\v) \\
          \end{array}
        \right] \succeq \mathbf{0}.  \label{denoiseopt2}
\end{align}

The above algorithm can be efficiently implemented via utilizing CVX tool \cite{cvx}.
Meanwhile, the 2D-frequency components can be obtained via the same Vandermonde decomposing methods for Toeplitz matrices $\mathcal{T}(\u), \mathcal{T}(\v) $  that are dealt with in  (\ref{theta1D}) and   paired technique in (\ref{pairing}).

In order to guarantee the correctness of the denoising and estimation algorithm (\ref{denoiseopt2}), two useful propositions about Optimal solution  condition and error bound for (\ref{denoise}) are concluded as the following, with theirs proofs are given in Appendix B.
\begin{prop}
(Optimality Conditions) If $\hat{\X}$ is the solution of (\ref{denoise}) then it holds that
\begin{align}
\begin{cases}
\|\Y-\hat{\X}\|_{\mathcal{A}^{'}_m}  \leq  \lambda\\
\langle \Y-\hat{\X},\hat{\X} \rangle = \lambda \|\hat{\X}\|_{\mathcal{A}_m}.
\end{cases}  \label{nop1}
\end{align}
\end{prop}
\begin{prop} (Error Bound)
If $\|\W \|_{\mathcal{A}^{'}_m} \leq \lambda $, then the optimum solution $\hat{\X}$ of (\ref{denoise}) satisfies
\begin{align}
\|\hat{\X} - \X^\sharp\|_F^2 \leq  2\lambda \|\X^\sharp \|_{\mathcal{A}_m}.
\end{align}
\end{prop}

It is often the case that the observed noises are random noise, hence it can be obtained that expected error rate of estimation in (\ref{denoise}) is as following whose proof is detailed in Appendix B.
 \begin{Thm} \label{thm2}
 Assume the entries of $\W$ are i.i.d. Gaussian entries with $ w_{i,j} \sim \mathcal{CN}(0,\sigma^2)$,  let
 \begin{align}
\lambda =
2\sigma\sqrt{MN} \Big{(}& \sqrt{\ln[16\pi^2(M-1)(N-1)]} \notag  \\
&+\frac{128\pi^2 (M-1)^2(N-1)^2}{\sqrt{\ln[16\pi^2(M-1)(N-1)]}}\Big{)} \label{lamda}
\end{align}
then the expected mean square error of solution $\hat{\X}$ to (\ref{denoise}) is bounded as \beq
 \mathds{E}\{ \| \hat{\X}-\X^\sharp\|^2_F \} \leq  2\lambda \|\X^\sharp\|_{\mathcal{A}_m}.      \label{estimbound}
 \eeq
 \end{Thm}
Note that, in the above theorem $\lambda$ is proportional to the noise standard deviation $\sigma$, which implies the weaker the noise level the more precious the estimation accuracy of data recovery is.
\section{Numerical experiments} \label{sim}
In this section, three kinds of experiments are conducted to verify the efficiency and examine the performance of the proposed algorithms (\ref{optalg1}) and (\ref{denoiseopt2}).
\subsection{MANM Recovery from Partial noiseless observation (\ref{optalg1})}
In this experiment, the dimension of array is set as $M=N=10$, for these 2D-harmonic signals that included in the received  data matrix $\X$,  their  2D-frequency pair of $f_{x,i}$ and $f_{y,i}$  are randomly located in  $[0,1]\times [0,1]$ satisfy the separation condition $\Delta=\mathrm{min~max}_{i\neq k}\{ |f_{x,i}-f_{x,k}|,|f_{y,i}-f_{y,k}|\} > \frac{1.19}{M}$ as in\cite{Chi}, and  their  signal amplitude $\s \in \mathds{C}^K$ satisfy i.i.d. complex Gaussian distribution.
Each entry in $\X^\sharp$ was observed with equal probability of $p=\frac{|\Omega|}{MN}$ via sub-Nyquist sampling, where $\Omega$ is the index set of observation entries and $|\Omega|$ denotes the number of observation entries in $\X^\sharp$ .  All simulation for algorithm  (\ref{optalg1}) was implemented  using CVX \cite{cvx}. The performance of recovery algorithm (\ref{optalg1}) are evaluated by the normalized reconstruction error as $ \frac{\|\hat{\X}-\X^\sharp \|_F}{\|\X^\sharp\|_F}$, and the reconstruction is considered to be successful if its reconstruction error less than $10^{-4}$.

The successful recovery rates are obtained from average results of conducting a total of 300 trials, then successful recovery rate versus the number of observation entries $|\Omega|$ are shown in Fig. 2 and Fig. 3 with $K=2,4,6,8$ respectively. It can be seen that the more the observation entries are, the higher the success rate becomes for the same sparsity level. And the larger the number of 2D-frequency pairs is the lower the successful recovery rate turns for the same observation level $p$.
\begin{figure}[!htbp]
      \centering
       \includegraphics[width=7.2cm]{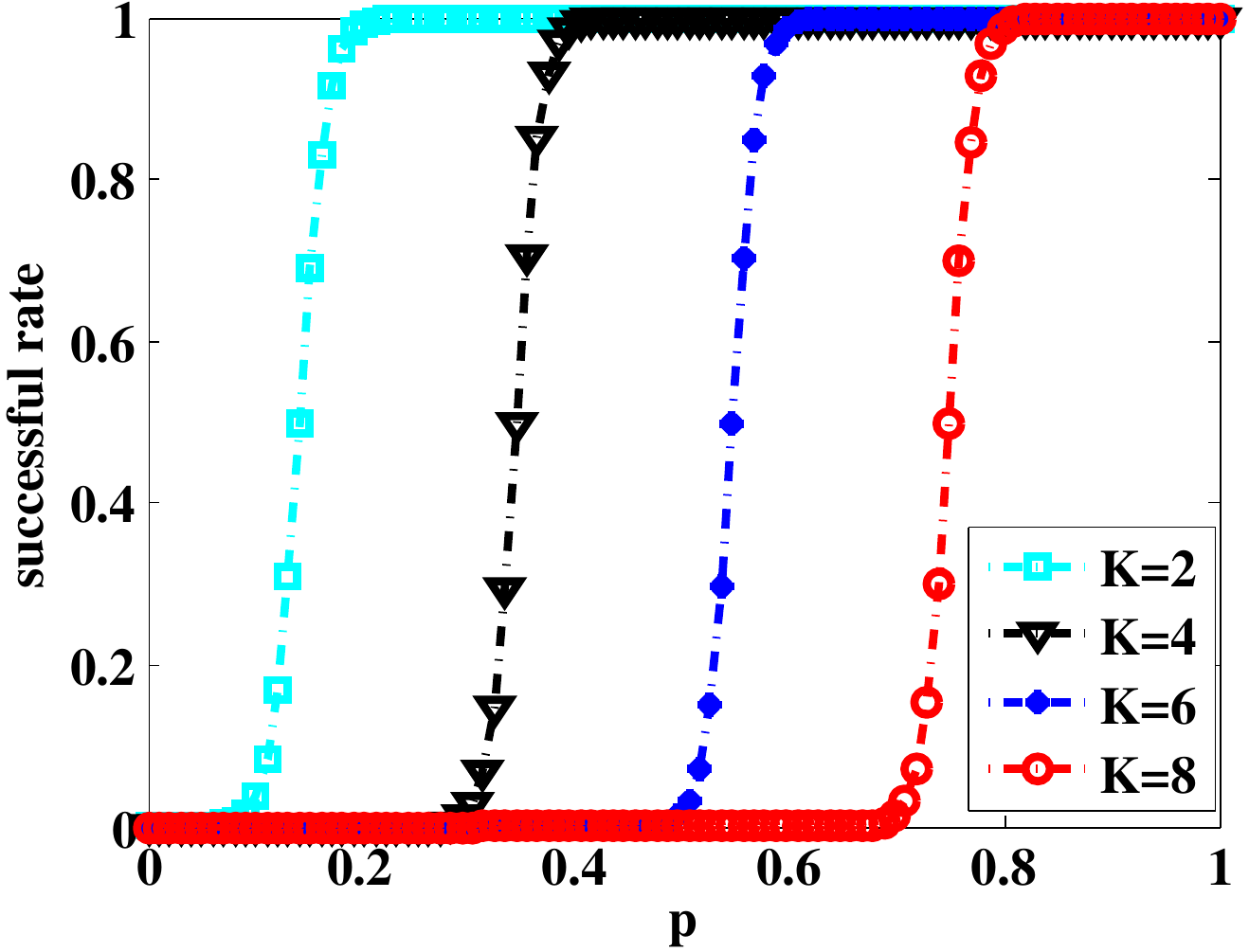}
    \caption{Successful rate of reconstruction versus efficient observation rate $p$ for $K=2,4,6,8$  when $M=N=10$.}
    \label{fig_d}
\end{figure}
\begin{figure}[!htbp]
      \centering
       \includegraphics[width=7.2cm]{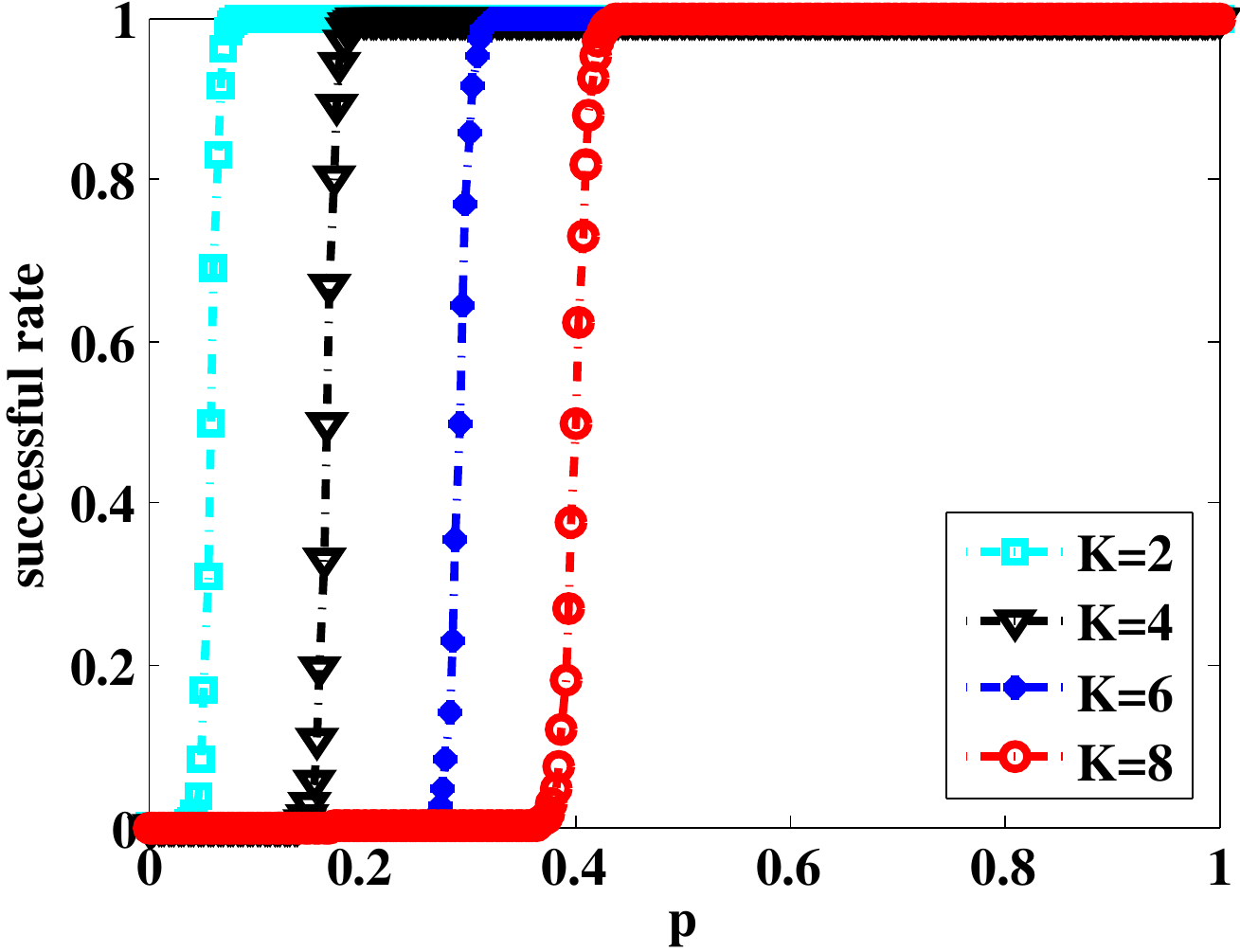}
    \caption{Successful rate of reconstruction versus efficient observation rate $p$ for $K=2,4,6,8$  when $M=N=18$.}
    \label{fig_d}
\end{figure}
\subsection{MANM denoise from full noise data  (\ref{denoiseopt2})}
In this experiment, the dimension of array is set as $ M=N=12$ and the number of 2D-harmonic signals is set to $K=6$ and their 2D-frequencies are randomly generated frequency pairs in $[0,1] \times [0,1]$ to satisfy the separation condition $\Delta > \frac{1.19}{M}$. Meanwhile the coefficient of each frequency was generated with constant magnitude one and a random phase from $[0,2\pi]$. The noise matrix $\W$ is randomly generated with i.i.d. complex Gaussian distribution, i.e., $w_{i,j}\sim \mathcal{CN}(0,\sigma^2)$. The vector mean square error (MSE) of the data matrix denoising estimation is defined as $\mathrm{MSE}(\hat{\X})\triangleq \mathds{E}\|\hat{\X}-\X^\sharp\|_F^2$ and obtained via taking average of 300 independent trials. The Signal-Noise-Ratio (SNR) is defined as $\mathrm{SNR} \triangleq \frac{1}{\sigma^2}$.
 \begin{figure}[!htbp]
      \centering
       \includegraphics[width=7.2cm]{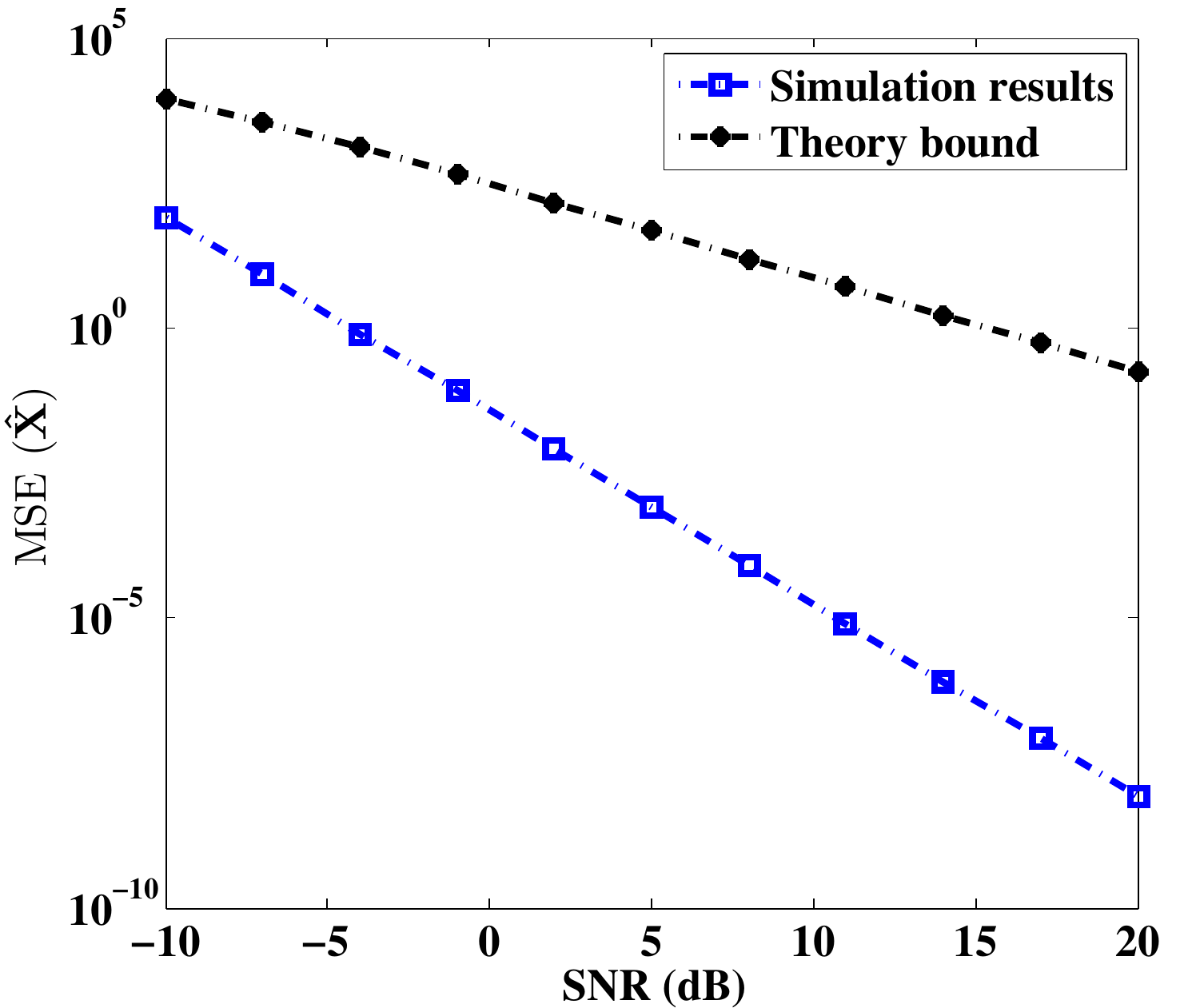}
    \caption{The $\mathrm{MSE}$ performance for data matrix $\X$ denoising via MANM method, and its theoretical upper bound, versus different $ \mathrm{SNR} $  level when $M=N=12, K=6$.}
    \label{fig_d}
\end{figure}

Fig. 4 shows the MSE performance of algorithm (\ref{denoise})  as well as
 the theoretical upper bound (\ref{estimbound}) obtained from Theorem 2, which can be seen that the MSE of data matrix denoising estimation decreases with the increasing of SNR.  Meanwhile,  the theoretical bound exhibits similar trends as the simulation performance, though  it's not sharp, it is an upper bound for data matrix estimation.

Furthermore,  the performance of 2D frequency component estimation of (\ref{thm2}) are evaluated with comparison against the Cramer-Rao-Bound (CRB). And the MSE of estimation is measured as
\beq
\mathrm{MSE}(\hat{\f})=\mathds{E}\|\hat{\f}-\f^\sharp \|^2=\mathds{E}\sum_{i=1}^K [(\hat{f}_{x,i}-f^\sharp_{x,i})^2 + (\hat{f}_{y,i}-f^\sharp_{y,i})^2]  \notag
\eeq
 where, $\hat{\f}$ and $\f^\sharp$ are the estimated 2D-frequency vector and the true 2D frequency vector respectively. And the simulated MSE of 2D-frequency estimation is obtained via averaging over 500 Monte Carlo runs with respect to the noise realization, while the CRB of 2D-frequency $\f^\sharp$ can be derived from the Fisher information matrix as is in \cite{MEMP}.

Fig. 5 shows the average MSE of 2D-frequency and the corresponding CRB
with respect to different SNR level. It can be seen that, with
the increase of SNR level, the average MSE of 2D-frequency estimation gradually approaches to 0.

 \begin{figure}[htbp]
      \centering
       \includegraphics[width=7.2cm]{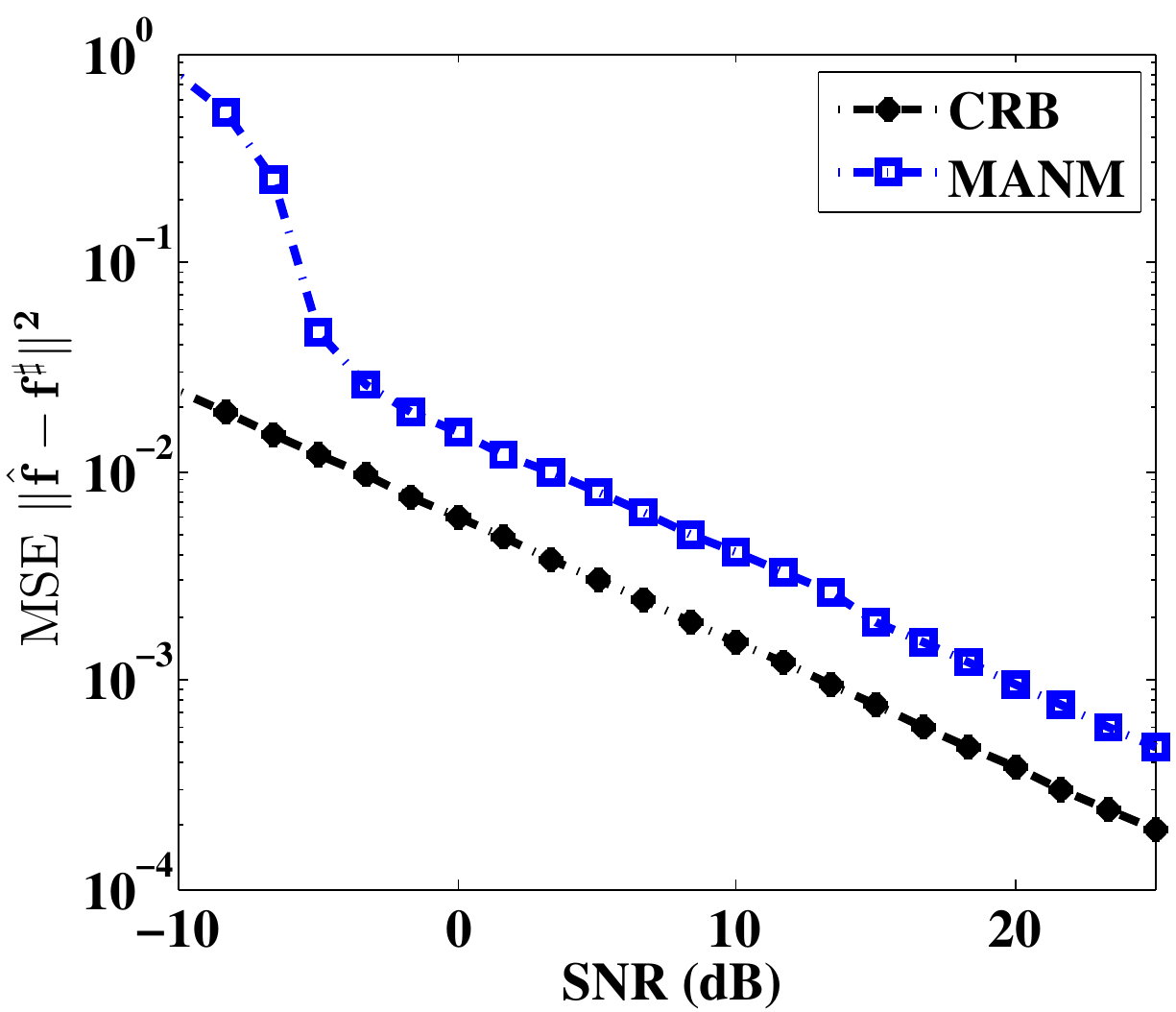}
    \caption{The $\mathrm{MSE}$ performance for 2D-frequency estimation via MANM method, and its $\mathrm{CRB}$, versus different $\mathrm{SNR}$ level when $M=N=12, K=6$.}
    \label{fig_d}
\end{figure}

\subsection{Comparisons With Existing Approaches}
In this experiment, the performance of proposed MANM method for 2D-frquency estimation  are compared with vectorial ANM \cite{Chi}  and other gridless sparse method such as basis pursuit (BP)  and orthogonal matched pursuit (OMP) method. As for  vectorial ANM method the VANM 2D-DOA is to solve
\beq
\hat{\X}= \underset{\X}{\mathrm{argmin}} \frac{1}{2}\|\Y-\X\|^2_F+\lambda\|\X\|_{\mathcal{A}_v} \label{VANMdenoise}
\eeq
where, $\|\X\|_{\mathcal{A}_v}$ is defined as in (\ref{VANM}) and the 2D-frquency are obtained via matrix enhancement matrix pencil (MEMP) method \cite{MEMP} .

 \begin{figure}[!t]
      \centering
       \includegraphics[width=7.2cm]{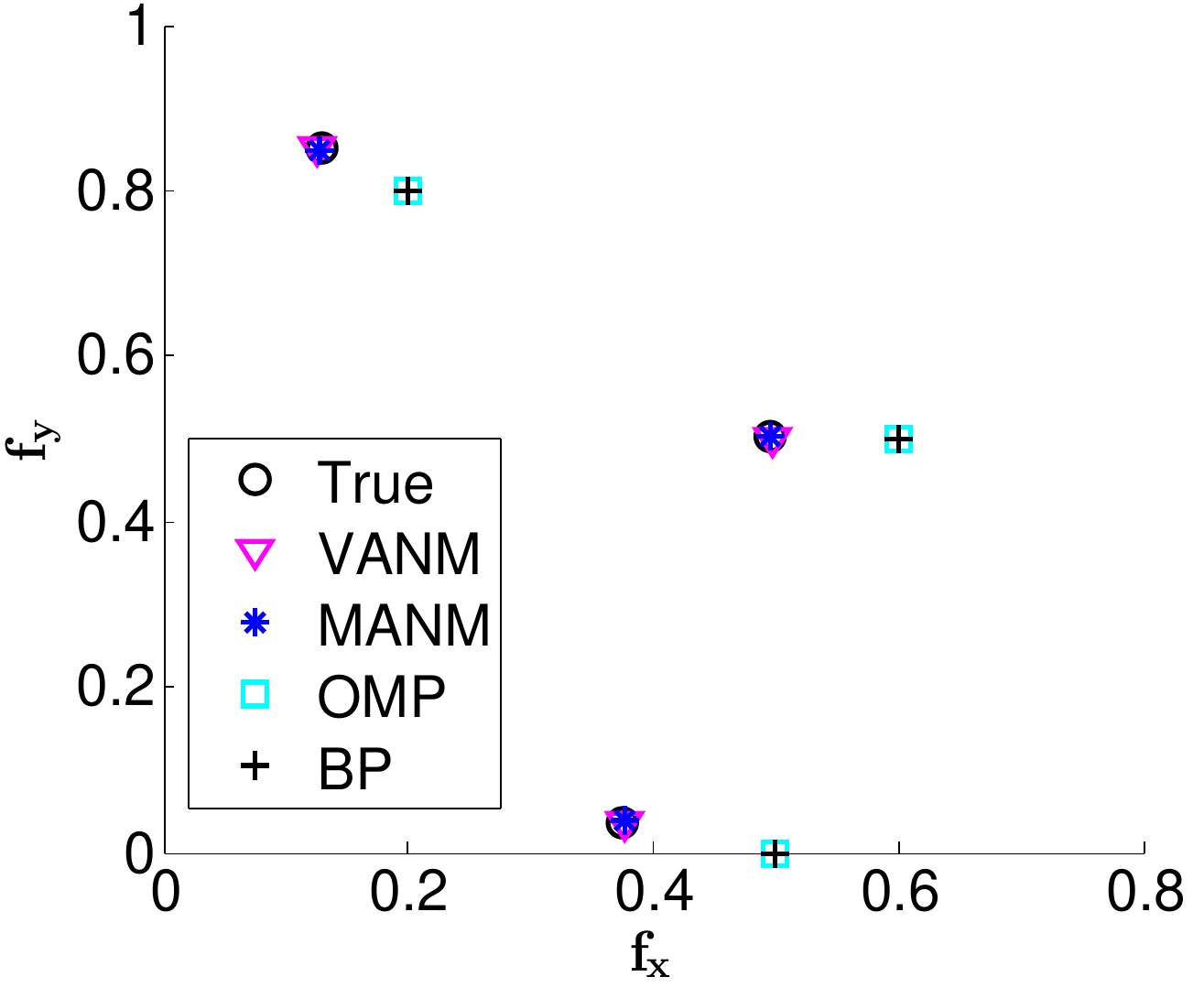}
    \caption{The 2D-frequencies estimation via different methods  ($M=N=10, K=3, \mathrm{SNR=6dB }, M_x=N_y=10$) }
    \label{fig_d1}
 \end{figure}

\begin{figure}[!t]
      \centering
       \includegraphics[width=7.2 cm] {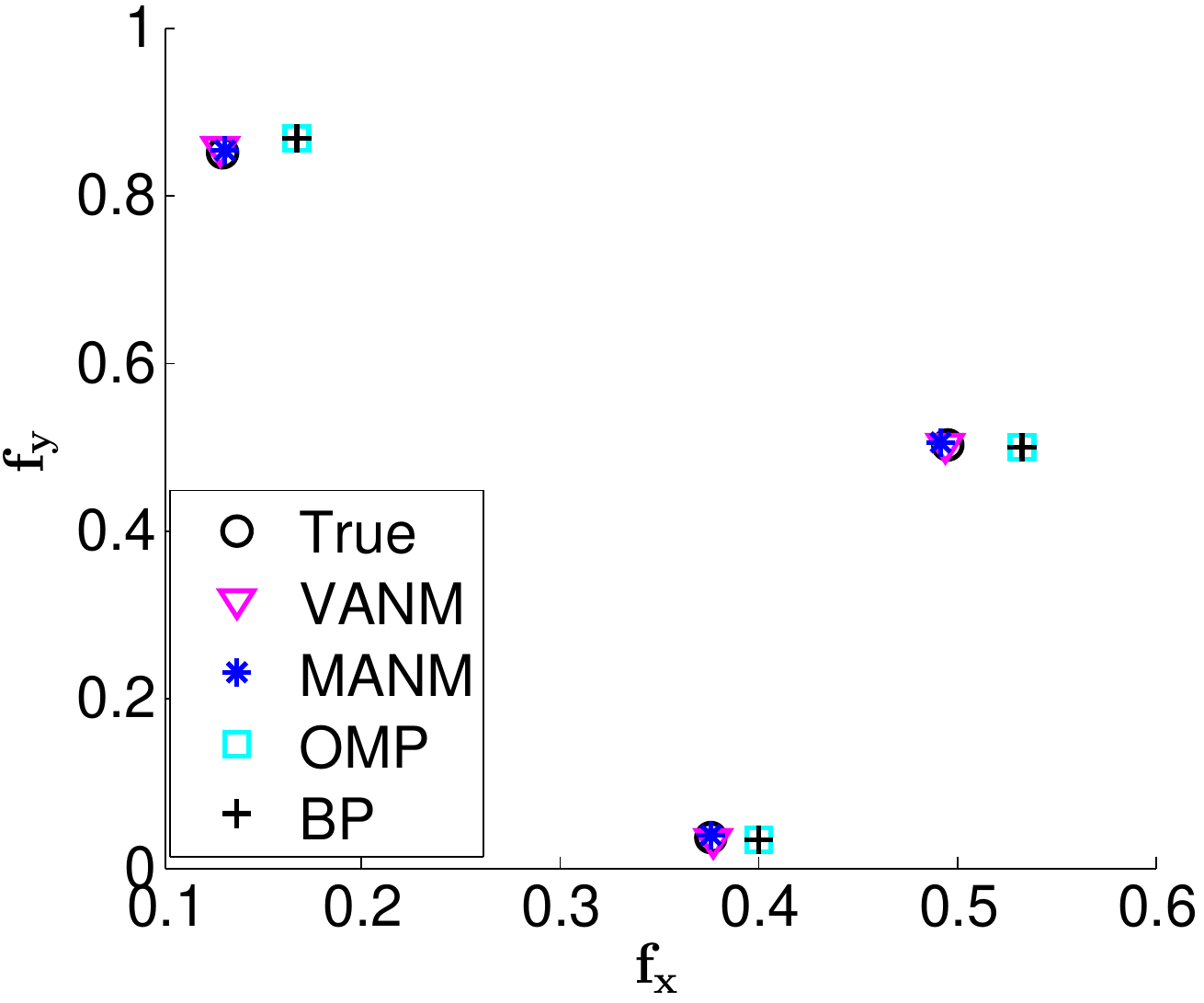}
    \caption{The 2D-frequencies estimation via different methods  ($M=N=10, K=3, \mathrm{SNR=6dB }, M_x=N_y=30$) }
    \label{fig_d2}
\end{figure}

 By split the 2D-frquency plane $[0,1]\times[0,1]$ into a large number of discrete grids, then grid  BP method can be applied to (\ref{noimod}) as
\beq
\underset{\s \in \mathds{C}^T} {\mathrm{argmin}}{\|\s\|_1} \quad s.t.\quad  \y=\PHi \s +\w  \label{L1}
\eeq
where, $ \PHi=\A_{_Y}(\f_{y}) \otimes \A_{_X}(\f_{x}),~\s=\vec {\S}~$and$~ \w=\vec{\W}$  are based on (\ref{Xdata2}). The interval $[0,1]\times[0,1]$ is divided into $M_x \times N_y$ grids to implement BP method as well as OMP method.

In order to compare the $\mathrm{MSE}$ performance of 2D -frequencies estimation via different methods, the parameters are set as  $\mathrm{SNR=6dB }$, and the 2D-frequencies are randomly generated in $[0,1) \times [0,1)$, with
$\{f_{x,k},f_{y,k}\}_{k=1:3}=\{(0.49546,0.50402), (0.37560,0.00369), (0.12951,0.85163)\}$ where the coefficient of each frequency was generated with constant magnitude one and a random phase from $[0,2\pi)$.  Typical 2D-frequencies estimation result for different methods are shown in Fig. 6 and Fig. 7, with $M_x=N_y=10$ and $M_x=N_y=30$, respectively.  It can be seen that with the more refined the grids are the better the 2D-DOA estimation results are for OMP and BP method, and the gridless methods of VANM and MANM methods gain better estimation accuracy than those grids methods such OMP and BP even if the grids are refined.

Furthermore, the $\mathrm{MSE}$ performance for 2D-frequency estimation via different methods and its $\mathrm{CRB}$ versus $\mathrm{SNR}$ are shown in Fig. 6, with all parameters are set as is in Fig. 8.  The simulated MSE of are obtained via averaging over 500 Monte Carlo runs with respect to the noise realization, and are plotted together in Fig. 6.  It can be seen that the $\mathrm{MSE}$ of all method decrease as the $\mathrm{SNR}$ increase. Compared with other method the MANM estimation method outperform all other method, which implies that the 2D-Frequencies estimation precision is better than others.

 \begin{figure}[!htbp]
      \centering
        \includegraphics[width=7.2cm]{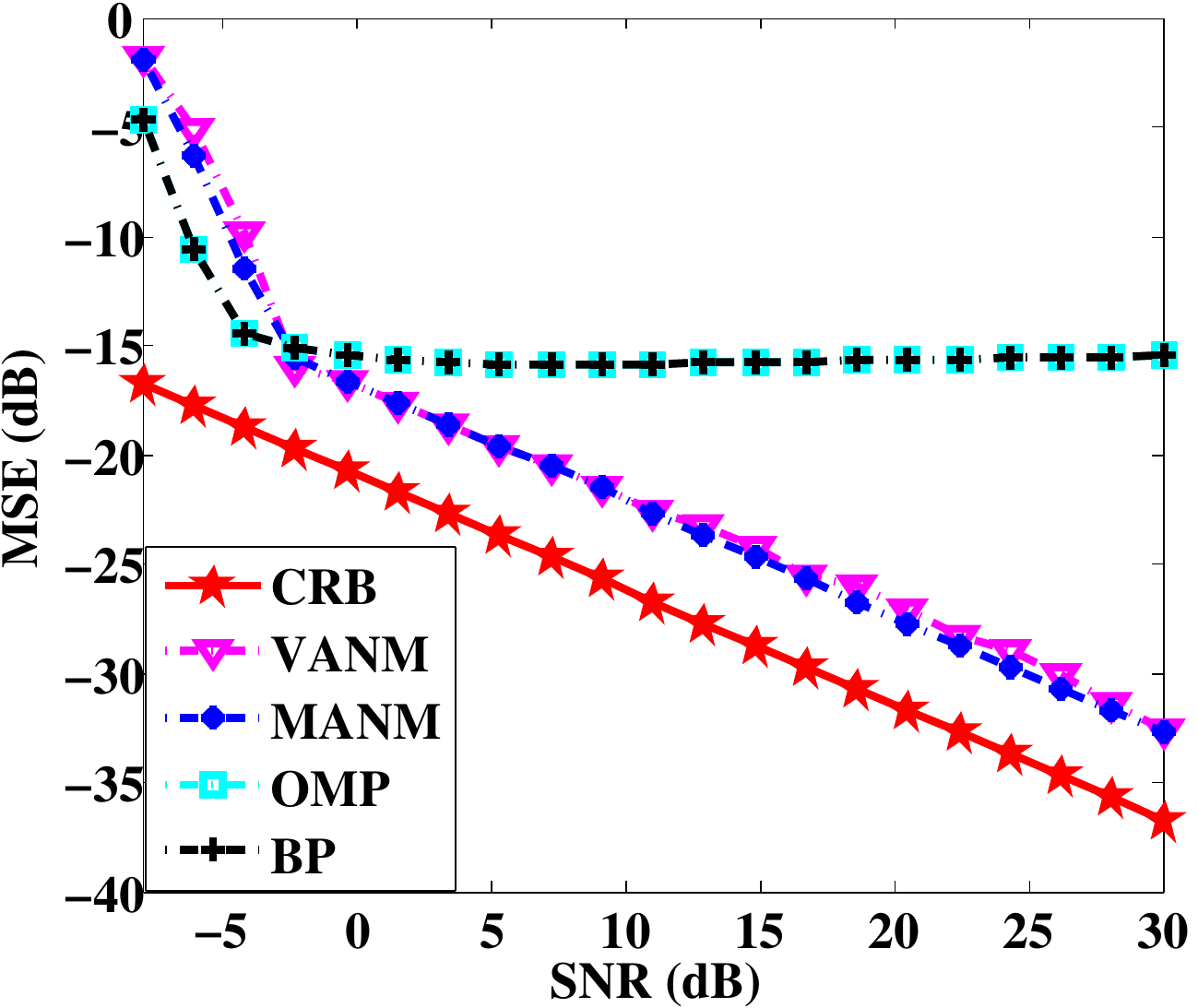}
     \caption{The $\mathrm{MSE}$ performance for 2D-frequency estimation via different methods  ($M=N=10, K=5, M_x=N_y=100$) }
     \label{fig_d}
 \end{figure}

 \begin{figure}[!htbp]
       \centering
       \includegraphics[width=7.2cm]{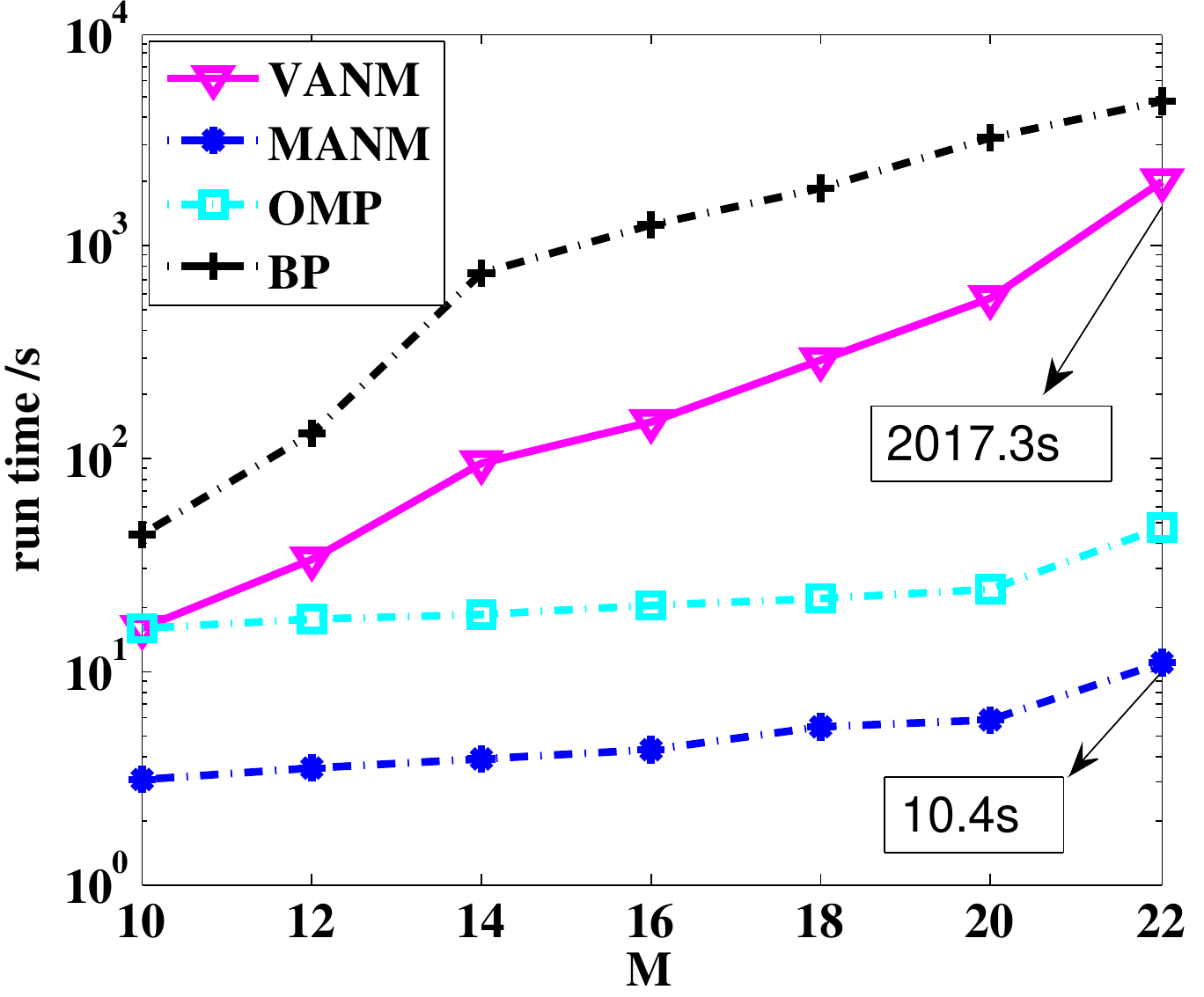}
     \caption{The run time for 2D-frequency estimation via different methods versus array dimension $M$ when   ($K=5,~~\mathrm{SNR=6dB }, M_x=N_y=100$) }
     \label{fig_d}
 \end{figure}

In the last experiment, the complexity of different algorithms are tested via their  running time versus dimension of array $M,N$ and are  shown in Fig. 7.  With the parameters are set as $M_x=N_y=100, K=5$ and $\mathrm{SNR=6dB }$, the experiment is conducted on a computer with Inter Core i7-3370 @3.4 GHz.  It can be seen from Fig. 7 that,  the time consuming of all methods increase with the increasing of  the dimension of array $M$ and $N$, the running time of VANM method increases much faster than the proposed MANM method.  When $M = N = 22$, the running time of the VANM is 2017.3s, while that of the MANM is only 10.4s. what's more, when the dimension of array is reach $M=N=35$ it takes almost a week to complete even one trial via VANM method, therefore, the MANM method earns a huge benefit in computational efficiency for large-scale arrays as well as retain high accuracy in 2D-Frequencies estimation.

\section{Conclusions}
In this paper, the problem of 2D-spectrum estimation and denoising  2D spectrally-sparse signals from their partial observations are studied, two approaches are developed and solved efficiently from matrix atomic norm minimizing via semi-definite programming other than vectorial atomic norm minimizing. The first algorithm aims to recover the full signal matrix from its partial observation, and the second algorithm aims to recover the structured signal matrix from full noise signal matrix. Theoretical performance guarantees are derived for both approaches under different conditions. Plenty of experiments are conducted to demonstrate  the efficiency of the proposed method, which greatly reduces the time spent on problem solving in large-scale array compared with the vectorial atomic norm minimizing, and retain the benefit of high estimation precision of atomic norm minimizing in 2D-Frequencies estimation.
\section{Appendix A \\ useful lemmas }
In this section we introduce a few useful lemmas, that will be used in our proofs.

\begin{lem}\label{lemvand}(Carthedory-toeplitz \cite{Carath1},\cite{Carath2})
Any positive semidefinite Toeplitz matrix $\T \in \mathds{C}^{N \times N}$with $\mathbf{rank}(\T)=r<N$, can be uniquely decomposed as
\begin{align}
\T& =\V\D\V^H  \notag\\
\V&=\left[
   \begin{array}{cccc}
     1 & 1 & ... & 1 \\
     v_1 & v_2 & ...& v_r \\
     : & : & ... & : \\
     v_1^N & v_2^N & ...& v_r^N
   \end{array}
   \right] \notag \\
   \D &=\mathbf{diag}(d_1,d_2,...,d_r) \notag \\
   \text{where} \quad &\quad
   \begin{cases}
    d_i>0, |v_i|=1\\
    v_i \neq v_j, \forall i\neq j
    \end{cases}  \quad i=1,2,...,r. \notag
\end{align}
\end{lem}

\begin{lem}\label{lem2}(Bernstein complex inequality \cite{bernstain1}) Let $P(z)=\sum_{k=0}^n p_kz^k$ be a polynomial with $deg(P(z))=n, ~p_k \in \mathds{C}$, then
 \beq
 \underset{|z|\leq 1}{\mathrm{sup}} |P^{'}(z)| \leq  n \underset{|z|\leq 1}{\mathrm{sup}} |P(z)|  \label{ bernstein}
 \eeq
\end{lem}

\begin{lem}\label{lem3}(Subdifferential of norm fuction \cite{sub1}\cite{sub2})
$\forall \x \in\mathds{X} \subseteq \mathds{C}^N $, the subdifferential of $\|\x\|_{_{ \mathds{X}}}$ is
\begin{align}
\partial\|\x_0\|_{_{ \mathds{X}}}=
\begin{cases}
\big{\{} \g \big{|} \g^H\x_0=\|\x_0\|_{_{ \mathds{X}}},\|\g\|_{_{ \mathds{X}^{'}}}=1, \g\in \mathds{C}^N \big{\}}    ~ \x_0 \neq \mathbf{0}\\
\big{\{} \g \big{|} \|\g\|_{_{ \mathds{X}^{'}}} \leq 1, \g\in \mathds{C}^N \big{\}}    \quad\quad\quad\quad\quad\quad~~~ \x_0 = \mathbf{0}.
\end{cases} \notag
\end{align}
where $(\mathds{X}^{'},\|\cdot\|_{_{\mathds{X}^{'}}})$ is the dual normal space of $(\mathds{X},\|\cdot\|_{_\mathds{X}})$.
\end{lem}

\blem \label{polypos1}(Bounded polynomials  \cite{pospoly})\\
Let $H(w)=\sum_{k=0}^{K-1}h_ke^{-jkw}$, $\h=[h_0,h_1,...,h_{K-1}]^T$, then
\bal
|H(w)|\leq \gamma, \quad \forall w \in [0,2\pi] \label{lem40}
\eal
holds, \textbf{iff}
\bal
       \begin{cases}
       \exists \quad  \Q \in \mathds{C}^{K\times K}, ~\Q=\Q^H  \\
       \gamma^2 \delta_k=\mathrm{Tr}(\THeta_{k}\Q), \quad k=0:K-1\\
       \left[
         \begin{array}{cc}
           \Q & \h\\
           \h^H & \Q\\
         \end{array}
       \right] \succeq \mathbf{0}\\
       \end{cases}
\eal
where, $\THeta_{k}\in \mathds{C}^{K\times K}$ is the elementary Toeplitz matrix with ones on the $k$-th diagonal and zeros elsewhere. i.e.,
        \bal
       &~|------\rightarrow k \notag \\
       \THeta_{k}=& \left[
         \begin{array}{cccccccc}
           0 & \cdots & 0& 1 & 0 &\cdots & \cdots & 0 \\
           0 & 0& \cdots & 0 & 1&0& \cdots & 0 \\
           \vdots & \ddots & \ddots& \cdots & 0& \ddots  & \cdots& 0 \\
           \vdots & \vdots & \ddots & \ddots & \cdots &\cdots & 1& 0 \\
           \mathbf{0} & \mathbf{0} & \mathbf{0} & \mathbf{0} & \mathbf{0}&\ddots& \mathbf{0} & 1 \\
           \vdots & \ddots & \ddots & \ddots & \ddots&\ddots& \ddots  &\mathbf{0} \\
         \end{array}
       \right] \label{etoep}
       \eal
\elem

\blem \label{polypos2}(Bounded polynomials with matrix coefficients  \cite{pospoly})
Let $\H_a(\theta)=\H \a(\theta), \a(\theta)=[1,e^{-j\theta},...,e^{-j(K-1)\theta}]^T$, then
\bal
\sIgma_{max}(\H_a(\theta)) \leq \gamma,\quad \forall  \theta \in [0,2\pi] \label{lem50}
\eal
holds, \textbf{iff}
\bal
       \begin{cases}
       \exists \quad  \Q \in \mathds{C}^{K\times K}, ~\Q=\Q^H  \\
       \gamma^2 \delta_k=\mathrm{Tr}(\THeta_{k}\Q), \quad k=0:K-1\\
       \left[
         \begin{array}{cc}
           \Q & \H\\
           \H^H & \I_K\\
         \end{array}
       \right] \succeq \mathbf{0}\\
       \end{cases} \label{pp}
\eal
where, $\THeta_{k}\in \mathds{C}^{K\times K}$ is the elementary Toeplitz matrix with ones on the $k$-th diagonal and zeros elsewhere, which is same as in (\ref{etoep}).
\elem
\section{Appendix B \\ proof of theorem  \ref{Thm1}}
\subsection{equivalent condition of dual norm}
\begin{prop}\label{pop4}
Let $\Z \in \mathds{C}^{M\times N}$ and $\|\Z\|_{\mathcal{A}^{'}_m}$ be the dual norm of matrix atomic norm that defined as in (\ref{dual1}) then
\bal
     & \|\Z\|_{\mathcal{A}^{'}_m} \leq 1 \quad ~~~  \Leftrightarrow \notag \\
     & \begin{cases}
       \exists \quad \P \in \mathds{C}^{N\times N}, \Q \in \mathds{C}^{M\times M}, ~ \P=\P^H, \Q=\Q^H  \\
       \delta_n=\mathrm{Tr}(\THeta^{_{(1)}}_{n}\P), \quad n=0:N-1\\
       \delta_m=\mathrm{Tr}(\THeta^{_{(2)}}_{m}\Q), \quad m=0:M-1\\
       \left[
         \begin{array}{cc}
           \Q & \Z\\
           \Z^H & \P \\
         \end{array}
       \right] \succeq \mathbf{0}\\
       \end{cases}
       \eal
       where, $\THeta^{_{(1)}}_{n} \in \mathds{C}^{N\times N}$ and $\THeta^{_{(2)}}_{m} \in \mathds{C}^{M\times M}$ are the elementary Toeplitz matrix with ones on the $n$-th diagonal and ones on the $m$-th diagonal, respectively, which are similar to (\ref{etoep}).
\end{prop}
\bproof
    Let
    \bal
    \begin{cases}
    \h(f_{x})=\Z \a_{{_X}(f_{x})},\h=\h(f_{x})\\
     H(f_{y})=\h^H\cdot \a_{{_Y}(f_{y})}=\sum_{k=0}^{N-1}h_kz^{-k}|_{z=e^{j\cdot f_{y}}}
     \end{cases}
     \label{eq1}
      \eal
      then according to lemma \ref{polypos1}, it can be obtained that
      \bal
      \| \Z\|_{\mathcal{A}^{'}_m}&\leq 1 \Leftrightarrow |H(f_{y})|\leq 1, \forall f_{y},f_{x}\in[0,2\pi)\\
       & \Leftrightarrow
       \begin{cases}
       \exists \quad \P \in \mathds{C}^{M\times M},~\P=\P^H  \\
       \delta_n=\mathrm{Tr}(\THeta^{_{(1)}}_{n}\P), \quad n=0:N-1\\
       \left[
         \begin{array}{cc}
           \P & \h \\
           \h^H& 1 \\
         \end{array}
       \right] \succeq \mathbf{0}  ~~~\forall f_x \in [0,2\pi]
       \end{cases} \label{pr4a}
       \eal
       where, $\THeta^{_{(1)}}_{n}\in \mathds{C}^{N\times N}$ is the elementary Toeplitz matrix with ones on the $n$-th diagonal and zeros elsewhere. Since
     \bal
     \left[
         \begin{array}{cc}
           \P & \h \\
           \h^H& 1 \\
         \end{array}
       \right]\succeq \mathbf{0}
       \Leftrightarrow
       1-\h^H\P^{-1}\h \geq 0  \notag \\
       \Leftrightarrow \a_{{_X}(f_{x})}^H \Z^H \P^{-1}\Z\cdot \a_{{_X}(f_{x})} \leq 1 \label{poseq1}
     \eal
     Let $\b(f_{x})\triangleq \P^{-\frac{1}{2}}\Z \cdot \a_{{_X}(f_{x})}$, then (\ref{poseq1}) is equivalent to
     \bal
     \b\b^H \preceq \I_M \Leftrightarrow\sIgma_{\max}{(\b(f))} \leq 1, \forall f\in [0,2\pi] \label{poseq2}
     \eal
      Let $\H_b\triangleq \P^{-\frac{1}{2}}\Z$, then according to lemma \ref{polypos2}, (\ref{poseq2}) is identical to
     \bal
     \begin{cases}
       \exists \quad \Q \in \mathds{C}^{M\times M}, ~ \Q=\Q^H  \\
       \delta_m=\mathrm{Tr}(\THeta^{_{(2)}}_{m}\Q), \quad m=0:M-1\\
       \left[
         \begin{array}{cc}
           \Q & \H_b \\
           \H_b^H& \I_N \\
         \end{array}
       \right] \succeq \mathbf{0}\\
       \end{cases} \label{pr4b}
       \eal
       where, $\THeta^{_{(2)}}_{m}\in \mathds{C}^{M\times M}$ is the elementary Toeplitz matrix with ones on the $m$-th diagonal and zeros elsewhere. then it comes
       \bal
       \Q-\H_b\H_b^H \succeq \mathbf{0} & \Leftrightarrow \Q-\Z^H\P^{-1}\Z \succeq \mathbf{0} \notag \\
       & \Leftrightarrow
       \left[
         \begin{array}{cc}
           \Q & \Z \\
           \Z^H & \P \\
         \end{array}
       \right] \succeq \mathbf{0} \label{pr4c}
       \eal
       based on (\ref{pr4a})(\ref{pr4b})(\ref{pr4c}), it can be derived that
       \bal
       &\|\Z\|_{\mathcal{A}^{'}_m} \leq 1 ~~\Leftrightarrow~~\notag \\
       & \begin{cases}
       \exists \quad \P \in \mathds{C}^{N\times N}, \Q \in \mathds{C}^{M\times M}, ~ \P=\P^H, \Q=\Q^H  \\
       \delta_n=\mathrm{Tr}(\THeta^{_{(1)}}_{n}\P), \quad n=0:N-1\\
       \delta_m=\mathrm{Tr}(\THeta^{_{(2)}}_{m}\Q), \quad m=0:M-1\\
       \left[
         \begin{array}{cc}
           \Q & \Z\\
           \Z^H & \P \\
         \end{array}
       \right] \succeq \mathbf{0}\\
       \end{cases}
       \eal
\eproof

\subsection{proof of theorem 1}
\bproof
According to the definition it holds that
\bal
\|\X\|_{\mathcal{A}_m}=\underset{\|\Z\|_{\mathcal{A}^{'}_m}\leq 1}{\sup} |\mathrm{Tr}(\X^H\Z)|=\underset{\|\Z\|_{\mathcal{A}^{'}_m}\leq 1}{\sup} \mathrm{Re}[\mathrm{Tr}(\X^H\Z)] \label{anm}
\eal
then according to proposition 4 it can be obtained that
 \bal
 &
 \|\X\|_{\mathcal{A}_m} =\sup ~\mathrm{Re} [\mathrm{Tr}(\X^H\Z)] \notag \\
 & s.t.\begin{cases}
       \exists \quad \P \in \mathds{C}^{N\times N}, \Q \in \mathds{C}^{M\times M}, ~ \P=\P^H, \Q=\Q^H  \\
       \delta_n=\mathrm{Tr}(\THeta^{_{(1)}}_{n}\P), \quad n=0:N-1\\
       \delta_m=\mathrm{Tr}(\THeta^{_{(2)}}_{m}\Q), \quad m=0:M-1\\
       \left[
         \begin{array}{cc}
           \Q & \Z\\
           \Z^H & \P \\
         \end{array}
       \right] \succeq \mathbf{0}\\
  \end{cases} \label{opt1}
 \eal
 the Lagrange augment function of (\ref{opt1}) is
\bal
L&(\Q,\P,\Y, \Z,\T_1,\T_2) \notag \\
&= -\frac{1}{2}\mathrm{Tr}(\X^H\Z+\Z^H\X)-\sum_{n=0}^{N}\mu_n\left(\delta_n-\mathrm{Tr}(\THeta^{_{(1)}}_{n}\P)\right)\notag \\
& -\sum_{m=0}^{M}\eta_m\left(\delta_m-\mathrm{Tr}(\THeta^{_{(2)}}_{m}\Q)\right)\notag \\
&+ \mathrm{Tr} \left(
\left[
  \begin{array}{cc}
    \T_1 &\Y \\
    \Y^H &\T_2  \\
  \end{array}
\right]
 \left[
     \begin{array}{cc}
       \Q & \Z\\
       \Z^H & \P \\
     \end{array}
 \right]
\right)
\eal
where $\left[
  \begin{array}{cc}
    \T_1 &\Y \\
    \Y^H &\T_2  \\
  \end{array}
\right]\succeq \mathbf{0} $. Then $\|\X\|_{\mathcal{A}_m} = - \inf L(\cdot) $ and the optimum value satisfy
\bal
\begin{cases}
\frac{\partial L}{\partial \Q}= \mathbf{0} \\
\frac{\partial L}{\partial \P}= \mathbf{0}\\
\frac{\partial L}{\partial \Z}= \mathbf{0}
\end{cases}
 \Leftrightarrow
 \begin{cases}
 \T_1=\sum_{m=0}^{M}\eta_m\THeta^{_{(2)}}_{m}\\
 \T_2=\sum_{n=0}^{M}\mu_n\THeta^{_{(1)}}_{n}\\
 \Y=\frac{1}{2}\X
\end{cases} \label{optcon1}\\
\inf (L)= \underset{\T_1,\T_2} {\inf} [-(\eta_0+\mu_0)]   \label{optcon2}
\eal
according to the definition of $\THeta^{_{(2)}}_{m}$ and $\THeta^{_{(1)}}_{n}$, it can be obtained that (\ref{optcon1}) imply $\T_1$ and $\T_2$ are both Hermite positive Toeplitz matrices with vectors $\eTa=[\eta_0,\eta_1,...,\eta_{M-1}]^T$ and $\mU=[\mu_0,\mu_1,...,\mu_{N-1}]^T$ as their first column respectively. Then it holds
\bal
\inf (L)= &\underset{\T_1,\T_2} {\inf} [-(\eta_0+\mu_0)] \notag \\   =&\underset{(\eTa,\mU) \in\mathds{C}^M \times \mathds{C}^N} {\inf}~ \mathrm{Tr}[ -\frac{1}{M} \mathcal{T}(\eTa) - \frac{1}{N} \mathcal{T}(\mU)] \label{optcon3}
\eal
Therefore it can be obtained that
\bal
\|\X\|_{\mathcal{A}_m}&=\underset{(\eTa,\mU) \in\mathds{C}^M \times \mathds{C}^N} {\inf}~ \mathrm{Tr}[ \frac{1}{M} \mathcal{T}(\eTa)+ \frac{1}{N} \mathcal{T}(\mU)] \notag \\
& s.t.
\left[
  \begin{array}{cc}
    \mathcal{T}(\eTa) &\frac{1}{2}\X \\
    \frac{1}{2}\X^H &\mathcal{T}(\mU)
  \end{array}
\right]\succeq \mathbf{0}
\eal
which is equivalent to
\bal
\|\X\|_{\mathcal{A}_m}&=\underset{(\u,\v) \in\mathds{C}^M \times \mathds{C}^N} {\inf}~ \mathrm{Tr}[ \frac{1}{2M} \mathcal{T}(\u)+ \frac{1}{2N} \mathcal{T}(\v)] \notag \\
& s.t.
\left[
  \begin{array}{cc}
    \mathcal{T}(\u) &\X \\
    \X^H &\mathcal{T}(\v)
  \end{array}
\right]\succeq \mathbf{0}
\eal
\eproof
\section{Appendix C \\ proof of proposition 2,3 and theorem 2}

\subsection{proof of proposition 2}
\begin{IEEEproof}
Let $h(\X)=\frac{1}{2}\|\Y-\X\|^2_F+\lambda\|\X\|_{\mathcal{A}_m}$, then it can be obtained that
\beq
\begin{cases}
\partial h( \X )=(\hat{\X}-\Y)^{*}+ \lambda \Z \\
\Z \in  \partial \|\X\|_{\mathcal{A}_m}
\end{cases}
\eeq
Since $\hat{\X}$ is the solution of (\ref{denoise}) then it holds that
\beq
\mathbf{0}\in \partial h(\hat{\X}) \label{pr3a1}
\eeq
according to lemma (\ref{lem3}) it has
\begin{align}
\begin{cases}
(\Y-\hat{\X})^{*} = \lambda \Z \\
\langle \hat{\X},\Z \rangle =  \|\hat{\X}\|_{\mathcal{A}_m}\\
 \|\Z\|_{\mathcal{A}^{'}_m} \leq 1
\end{cases} \notag
\end{align}
then it turns out
\begin{align}
\begin{cases}
\|\Y-\hat{\X}\|_{\mathcal{A}^{'}_m}  \leq  \lambda\\
\langle \Y-\hat{\X},\hat{\X} \rangle =\lambda \|\hat{\X}\|_{\mathcal{A}_m}
\end{cases} \notag
\end{align}
\end{IEEEproof}

\subsection{proof of proposition 3}
\begin{IEEEproof}
since $\Y=\X^\sharp+\W $ then
\begin{align}
\|\hat{\X} - \X^\sharp\|_F^2&=\langle \hat{\X} - \X^\sharp, \W-(\Y-\hat{\X}) \rangle \notag \\
&=
\langle \hat{\X} - \X^\sharp, \W-(\Y-\hat{\X}) \rangle _{\mathbf{R}}
\notag \\
&=
\langle \X^\sharp,\Y-\hat{\X}\rangle _{\mathbf{R}} - \langle \X^\sharp,\W \rangle _{\mathbf{R}}
 \notag \\
&\quad +   \langle \hat{\X} ,\W \rangle _{\mathbf{R}} -    \langle \hat{\X},\Y-\hat{\X}\rangle _{\mathbf{R}}
\label{stp2} \\
& \leq \|\Y-\hat{\X}\|_{\mathcal{A}^{'}_m}\|\X^\sharp\|_{\mathcal{A}_m} +
\|\W\|_{\mathcal{A}^{'}_m} \|\X^\sharp\|_{\mathcal{A}_m}
\notag \\
&\quad +
\|\W\|_{\mathcal{A}^{'}_m}\|\hat{\X}\|_{\mathcal{A}_m} - \lambda\|\hat{\X}\|_{\mathcal{A}_m}
\label{stp3} \\
& \leq 2\lambda\|\X^\sharp\|_{\mathcal{A}_m} + (\|\W\|_{\mathcal{A}^{'}_m}-\lambda)\|\hat{\X}\|_{\mathcal{A}_m}
\notag \\
& \leq 2\lambda\|\X^\sharp\|_{\mathcal{A}_m}  \label{stp4}
\end{align}
where,  (\ref{stp2}) results from (\ref{dual2}),   (\ref{stp3}) results from (\ref{nop1}), (\ref{stp4}) results from  $\|\W \|_{\mathcal{A}^{'}_m} \leq \lambda $.
\end{IEEEproof}

\subsection{proof of theorem 2}
\begin{IEEEproof}
here we utilize the bounded property of the expectation of dual norm of $\|\W\|_{\mathcal{A}_m}$ to prove the claim. According to the definition of dual norm, one has
\begin{align}
\|\W\|_{\mathcal{A}^{'}_d} &=\underset{\f \in[0,1]^2}{\mathrm{sup}}|\langle\W,\A_d(\f)\rangle| =\underset{(f_x,f_y) \in[0,1]^2}{\mathrm{sup}}|W(f_x,f_y)|
\end{align}
where
 \begin{align}
 W(f_x,f_y)&\triangleq \sum_{m=1}^M\sum_{n=1}^N w^*_{m,n} e^{-j2\pi[(m-1)f_x+(n-1)f_y]} \label{w1}
 \end{align}
let $(u,v)=(e^{-j2\pi f_x},e^{-j2\pi f_y})$ then
 \beq
  W(u,v)= \sum_{m=1}^M\sum_{n=1}^N w^*_{m,n} u^{m-1}v^{n-1} \quad \quad  |u| \leq 1,  |v| \leq 1
 \eeq
Note that $\forall u_1,u_2,v_1,v_2 \in \{\xi~|~|\xi|\leq 1 ,\xi\in \mathds{C}\}$, it has
\begin{align}
  |W_{(u_1,v_1)}-W_{(u_2,v_2)}| & \leq \|W_{u}^{'}\|_{_{\infty}}|u_1-u_2 | + \|W_{v}^{'}\|_{_{\infty}}|v_1-v_2| \label{uberst}
\end{align}
where
\begin{align}
 |u_1-u_2 | & \leq 2\pi |f_{1_x} - f_{2_x}|  \\
 |v_1-v_2 | & \leq 2\pi |f_{1_y} - f_{2_y}|  \\
 \|W_{u}^{'}\|_{_{\infty}} & \leq  (M-1)\|W \|_{_{\infty}} = (M-1)\|\W\|_{\mathcal{A}^{'}_m} \label{ber1}\\
 \|W_{v}^{'}\|_{_{\infty}} & \leq  (N-1)\|W \|_{_{\infty}}  = (N-1)\|\W\|_{\mathcal{A}^{'}_m}  \label{ber2}
\end{align}
where,  (\ref{ber1}) and (\ref{ber2}) follow by Bernstein's theorem \cite{bernstain1}, i.e. lemma \ref{lem2}. Then according to (\ref{w1}) to (\ref{ber2}), it has
\begin{align}
  |W_{(\f_1)}-W_{(\f_2)}| & \leq 2\pi(M-1) |f_{1_x} - f_{2_x}|\cdot\|\W\|_{\mathcal{A}^{'}_m} \notag \\
  &~   + 2\pi(N-1) |f_{1_y} - f_{2_y}| \cdot \|\W\|_{\mathcal{A}^{'}_m} \label{Adb1}
\end{align}
Letting $f_{2_x}\in \{0,\frac{1}{M_x},...,\frac{M_x-1}{M_x}\}, f_{2_y}\in \{0,\frac{1}{N_y},...,\frac{N_y-1}{N_y}\}$, it can be seen that
\begin{align}
\|\W\|_{\mathcal{A}^{'}_m} & \leq  \underset{m=0,..,M_x-1 \atop n=0,...,N_y-1}{\mathrm{max}}|W(e^{-j2\pi m/M_x},e^{-j2\pi n/N_y})| \notag \\
&\quad\quad+\pi\frac{M-1}{M_x} \|\W\|_{\mathcal{A}^{'}_m}+\pi\frac{N-1}{N_y} \|\W\|_{\mathcal{A}^{'}_m} \label{Adb3}
\end{align}
Therefore it has
\begin{align}
  \|\W\|_{\mathcal{A}^{'}_d} & \leq
  \Big{(} 1-\pi(\frac{M-1}{M_x}+\frac{N-1}{N_y})\Big{)} ^{-1} \notag \\
  &~~ ~\cdot \underset{m=0,..,M_x-1 \atop n=0,...,N_y-1}{\mathrm{max}}|W(e^{-j2\pi m/M_x},e^{-j2\pi n/N_y})|
\end{align}
let $ g_{m,n}=|W(e^{-j2\pi m/M_x},e^{-j2\pi n/N_y})|$,
since $w_{i,j}$ obey i.i.d. complex Gaussian distribution with $w_{i,j}\sim \mathcal{CN}(0,\sigma^2)$, then it can be obtained that
$ W(f_x,f_y) \sim \mathcal{CN}(0,MN\sigma^2) $ and $g_{m,n}$ satisfying  Rayleigh distribution
\beq
p(g)= \frac{2g}{MN\sigma^2} e^{-\frac{g^2}{MN\sigma^2}}  \quad  g \geq 0 \label{disg}
\eeq
which holds for $\forall 0\leq m \leq M_x, 0\leq n \leq N_y$, then we aim to find the minimum value of up bound of
$\mathds{E}(\underset{1\leq m \leq M_x-1 \atop 1 \leq n \leq N_y-1}{\mathrm{max}} g_{i,j} )$ by suitably choosing $M_x$ and $N_y$ as is dealt in \cite{Tang2}.

Therefore it can be seen that
\begin{align}
\mathds{E}(\underset{1\leq m \leq M_x-1 \atop 1 \leq n \leq N_y-1}{\mathrm{max}} g_{i,j} )& =\int_{0}^{+\infty} \mathds{P}(\underset{1\leq m \leq M_x-1 \atop 1 \leq n \leq N_y-1}{\mathrm{max}} g_{m,n} \geq t) dt \notag \\
& \leq \varepsilon +\int_{\epsilon}^{+\infty} \mathds{P}(\underset{1\leq m \leq M_x-1 \atop 1 \leq n \leq N_y-1}{\mathrm{max}} g_{m,n} \geq t) dt \notag \\
& \leq  \varepsilon + M_xN_y\int_{\varepsilon}^{+\infty} \mathds{P}( g_{0,0} \geq t) dt
\notag \\
& =  \varepsilon + M_xN_y\int_{\varepsilon}^{+\infty} e^{-\frac{t^2}{MN\sigma^2}} dt \label{Adb4}
\end{align}
Hence letting $\varepsilon=\sigma\sqrt{MN \ln(M_xN_y)}$, then
\beq
\mathds{E}(\underset{1\leq m \leq M_x-1 \atop 1 \leq n \leq N_y-1}{\mathrm{max}} g_{i,j} ) \leq
\sigma\sqrt{MN} \Big{(} \sqrt{\ln(M_xN_y)}+\frac{ M_x^2N_y^2}{2\sqrt{\ln(M_xN_y)}}\Big{)} \eeq
where, the second part of right hand side follows from inequality
\beq
\int_{u}^{+\infty}e^{-t^2/2} dt \leq \frac{1}{u}e^{-u^2/2}  \quad u >0
\eeq
then it can be obtained that
\begin{align}
\mathds{E}  \Big{\{}\|\W\|_{\mathcal{A}^{'}_m} \Big{\}}\leq
\frac{\sigma\sqrt{MN} \big{(} \sqrt{\ln(M_xN_y)}+\frac{ M_x^2N_y^2}{2\sqrt{\ln(M_xN_y)}}\big{)}
}{1-\pi(\frac{M-1}{M_x}+\frac{N-1}{N_y})}
\end{align}
let $M_x=4\pi(M-1), N_y=4\pi(N-1)$ then
\begin{align}
\mathds{E}  \Big{\{}\|\W\|_{\mathcal{A}^{'}_d} \Big{\}} \leq
2\sigma\sqrt{MN} \Big{(}& \sqrt{\ln[16\pi^2(M-1)(N-1)]}
\notag \\
&+\frac{128\pi^2 (M-1)^2(N-1)^2}{\sqrt{\ln[16\pi^2(M-1)(N-1)]}}\Big{)}
\end{align}
Let
\begin{align}
\lambda =
2\sigma\sqrt{MN} \Big{(}& \sqrt{\ln[16\pi^2(M-1)(N-1)]}
\notag \\
&+\frac{128\pi^2 (M-1)^2(N-1)^2}{\sqrt{\ln[16\pi^2(M-1)(N-1)]}}\Big{)} \notag
\end{align}
then, it can be seen that $\mathds{E}  \Big{\{}\|\W\|_{\mathcal{A}^{'}_m} \Big{\}} \leq \lambda$, according to proposition 3, it holds that
\begin{align}
 \mathds{E}\{ \| \hat{\X}-\X^\sharp\|^2_F \} \leq  2\lambda \|\X^\sharp\|_{\mathcal{A}_d}  \notag
 \end{align}
\end{IEEEproof}




\begin{thebibliography}{1} 
\bibitem{radar1}
M. I. Skolnik, ``Introduction to Radar System,"  3rd ed. New York: Mc-
Graw-Hill, 2002.

\bibitem{radar2}
J. Li and P. Stoica, ``MIMO radar with colocated antennas: Review of
some recent work," \emph{IEEE Signal Process. Mag.},  vol. 24, pp. 106¨C114,
Sep. 2007.

\bibitem{cm1}
A. M. Sayeed and B. Aazhang, ``Joint multipath-doppler diversity in
mobile wireless communications," \emph{IEEE Trans. Commun.}, no. 1, pp.
123¨C132, Jan. 1999.

\bibitem{cm2}
D. Tse and P. Viswanath, ``Fundamentals of Wireless Communication,"
Cambridge, U.K.: Cambridge Univ. Press, 2005.

\bibitem{img}
R. Willett, M. Duarte, M. Davenport, and R. Baraniuk, ``Sparsity and
structure in hyperspectral imaging: Sensing, reconstruction, and target
detection,"  \emph{IEEE Signal Processing Mag.}, vol. 31, no. 1, pp. 116¨C126,
Jan. 2014.

\bibitem{ESPRIT}
M. Haardt, M. Zoltowski, C. Mathews, J. Nossek, ``2D unitary ESPRIT for efficient 2D parameter estimation," \emph{Proc. IEEE Int. Conf. Acoust. Speech Signal Process}.   vol. 3, pp. 2096-2099, 1995.

\bibitem{MUSIC}
Y. Hua, ``A pencil-music algorithm for finding two-dimensional angles and polarizations using crossed dipoles,"  \emph{IEEE Trans. Antennas Propag.}, vol. 41, no. 3, pp. 370-376, 1993.

\bibitem{MEMP}
Y. Hua, ``Estimating two-dimensional frequencies by matrix enhancement and matrix pencil," \emph{IEEE Trans. Signal Process.}, vol. 40, no. 9, pp. 2267-2280, Sep. 1992.

\bibitem{coh}
H. L. Van Trees, ``Detection, Estimation, and Modulation Theory - Optimum Array Processing," (Part IV), Wiley, 2002.


\bibitem{CS}
 D.~Donoho, ``Compressed sensing,'' \emph{IEEE Trans. Inf. Theory.}, vol.~52, no.~4,pp.1289--1306, Apr. 2006.

\bibitem{CS2}
 E.~Candes, J.~Romberg, and T.~Tao, ``Robust uncertainty principles: Exact signal reconstruction from highly incomplete frequency information,'' \emph{IEEE Trans. Inf. Theory.}, vol.~52, no.~2, pp. 489-509, Feb. 2006.

\bibitem{CS3}
 M. Stojnic, ``$L_1$ optimization and its various thresholds in compressed sensing,"  \emph{ICASSP 2010,}  pp. 3910-3913.

\bibitem{BP}
 J. Tropp, ``Just relax: Convex programming methods for identifying sparse signals in noise", \emph{IEEE Trans. Inf. Theory.}, vol. 52, no. 3, pp. 1030-1051, 2006

\bibitem{LASSO}
  H. Zou,~``The adaptive lasso and its oracle properties," \emph{J. Amer. Statist. Assoc.}, vol. 101, no. 476, pp. 1418-1429, 2006.

\bibitem{OMP}
 J. A. Tropp, ``Greed is good: Algorithmicresults for sparse approximation", \emph{IEEE Trans.Inf. Theory.}, vol. 50, no. 10, pp. 2231-2242, 2004

\bibitem{PIP}
D. Ramasamy, S. Venkateswaran, U. Madhow, ``Compressive parameter estimation in AWGN," \emph{IEEE Trans. Signal Process.}, vol. 62, no. 8, pp. 2012-2027, Apr. 2014.

\bibitem{miss1}
Y. Chi, L. L. Scharf, A. Pezeshki, R. Calderbank, ``Sensitivity to basis mismatch in compressed sensing," \emph{IEEE Trans. Signal Process.}, vol. 59, no. 5, pp. 2182-2195, May. 2011.

\bibitem{miss2}
A. Fannjiang, H.-C. Tseng, ``Compressive radar with off-grid targets: A perturbation approach", Inverse Problems, vol. 29, no. 5, pp. 054008, 2013.

\bibitem{miss3}
Z. Tan, P. Yang, A. Nehorai, ``Joint sparse recovery method for compressed sensing with structured dictionary mismatch,"  \emph{IEEE Trans. Signal Process.},  vol.~62, no. 19, pp. 4997-5008, Oct. 2014.

\bibitem{Tang}
G. Tang, B. Bhaskar, P. Shah, B. Recht,  ``Compressed sensing off the grid,"  \emph{IEEE Trans. Inf. Theory.}, vol. 59, no. 11, pp. 7465-7490, 2013.

\bibitem{Carath1}
N. I. Ahiezer, Mark Grigor¨¹evich Krein, ``Some Questions in the Theory of Moments,'' \emph{Translations of Mathematical Monographs, American Mathematical Society.} 1985.

\bibitem{Carath2}
C. Carath¨¦odory and L. Fej¨¦r, ``Uber den zusammenghang der extemen von harmonischen funktionen mit ihren koeffizienten und uber den Picard-Landauschen satz," Rendiconti del Circolo Matematico diPalermo, vol. 32, pp. 218-239, 1911.

\bibitem{Tang2}
Badri Narayan Bhaskar, Gongguo Tang, and Benjamin Recht, ``Atomic norm denoising with applications to line spectral estimation," \emph{IEEE Trans. Signal Process.}, vol. 61, no. 23, pp. 5987-5999, 20l3.

\bibitem{super}
Emmanuel J Candes and Carlos Fernandez-Granda,
``Towards a mathematical theory of super-resolution,"
\emph{Communications on Pure and Applied Mathematics.},
vol. 67, no. 6,pp. 906-956,2014.

\bibitem{Chi}
Y. Chi, Y. Chen, ``Compressive two-dimensional harmonic retrieval via atomic norm minimization," \emph{IEEE Trans. Signal Process.}, vol. 63, no. 4, pp. 1030-1042, Feb. 2015.

\bibitem{VD}
Zai Yang, Lihua Xie, and Petre Stoica, ``Vandermonde
decomposition of multilevel Toeplitz matrices with application
to multidimensional super-resolution,"  \emph{IEEE
Transactions on Information Theory.}, vol. 62, no. 6, pp.
3685-3701, 2016.

\bibitem{Candes06}
 E.~Candes, J.~Romberg, and T.~Tao, ``Robust uncertainty principles: Exact signal reconstruction from highly incomplete frequency information,'' \emph{IEEE Trans. Inf. Theory.}, vol.~52, no.~2, pp. 489-509, Feb. 2006.

\bibitem{cvx}
 M. Grant and S.  Boyd, ``CVX: Matlab software for disciplined convex programming version 2.0 beta,'' [online]  Available: http://cvxr.com/cvx~Feb.~ 2015.

\bibitem{bernstain1}
A. Schaeffer, ``Inequalities of A. Markoff and S. Bernstein for polynomials
and related functions,¡± \emph{Bull.Amer. Math. Soc.}, vol. 47, pp.565¨C579, 1941.

\bibitem{bernstain2}
 R.P. Boas, JR, ``Inequalities for the derivatives of polynomials,¡±  \emph{Northwestern University, MATHEMATICS MAGAZINE}, Vol. 42, No. 4, September 1969.

\bibitem{sub1}
Rockafellar, R. Tyrrell, ``Convex Analysis,"  Princeton, \emph{ Princeton University Press.} (1997).

\bibitem{sub2}
Kusraev A. G. and Kutateladze S. S, ``Subdifferentials: theory and applications," \emph{Mathematics and its Applications} Vol. 323, Kluwer Academic Publishers, Dordrecht, 1995.

\bibitem{pospoly}
B. Dumitrescu, ``Positive Trigonometric Polynomials and Signal Processing Applications," New York, NY, USA: Springer, 2007.

\end{thebibliography}
\end{document}